\documentclass[%
 reprint,
 amsmath,amssymb,
 aps,
]{revtex4-2}

\usepackage{graphicx}
\usepackage{dcolumn}
\usepackage{bm}
\usepackage{xcolor}
\usepackage{textcomp}
\usepackage{float}
\usepackage{hyperref}

\usepackage{braket} 
\usepackage{caption}
\usepackage{subcaption}
\usepackage{booktabs}
\usepackage{algorithm}
\usepackage{multirow}
\usepackage[noend]{algpseudocode}

\DeclareCaptionFormat{myformat}{#3}
\begin{document}

\preprint{APS/123-QED}

\title{Physics-informed Kolmogorov–Arnold networks for
 viscoelastic fluid equations}

\author{Suryanshu Singh}
 \email{suryanshu22@iiserbpr.ac.in}
\affiliation{%
Department of Physics,  Indian Institute of Science Education and Research (IISER), Berhampur -760003, India 
}%
\author{Midhuna Suresh}
\email{midhuna000suresh@gmail.com}
\affiliation{Department of Physics, Maulana Azad National Institute of Technology (MANIT), Bhopal-462003, India}
\author{Akanksha Gupta }
\email{akanksha@manit.ac.in}
\affiliation{Department of Physics, Maulana Azad National Institute of Technology (MANIT), Bhopal-462003, India}

\date{\today}

\begin{abstract}
Kolmogorov–Arnold Networks (KANs), inspired by the Kolmogorov–Arnold representation theorem, provide an interpretable alternative to multilayer perceptrons (MLPs) by using learnable activation functions on edges rather than fixed node activations. We propose a Physics-Informed Kolmogorov–Arnold Network (PI-KAN) framework for solving forward problem of viscoelastic fluid equations, which arise in many complex fluid dynamics applications and are characterized by strong nonlinear coupling between fluid fields. For viscoelastic fluid equations, we adopt the generalized hydrodynamic model, which is well established in the field of dusty plasma. To evaluate the performance of the proposed framework for viscoelastic fluid, we consider benchmark problem based on the Taylor–Green (TG) flow and a modified Taylor–Green flow. We systematically investigate the effects of different network architectures, hyperparameters, and collocation point distributions on the accuracy and convergence behavior of PI-KANs for the range of viscoelastic parameter $(\tau_m=1-20)$. We also study the impact of  random seed initialization on training outcomes. The obtained results provide useful guidance for the design and implementation of physics-informed Kolmogorov–Arnold networks (PI-KANs) in solving viscoelastic fluid equations. 

\end{abstract}

\maketitle


\noindent

\section*{Introduction}
Physics-informed neural networks (PINNs) and related deep learning frameworks have demonstrated remarkable potential for solving partial differential equations (PDEs) \cite{Raissi2019_PINN, sirignano2018dgm, karniadakis2021physics, Raissi2018_HFM_arXiv, jagtap2020extended, Jagtap2020_cPINN, kharazmi2021hp, jin2021nsfnets, mahmoudabadbozchelou2022nn, thakur2024viscoelasticnet, lu2021deepxde, Yu2022GradBalance}. By employing governing physical laws, generally expressed as partial differential equations into the loss function, PINNs help us understand the hidden states and parameters from limited data while preserving physical consistency. 
This method has found widespread success in modeling fluid-mechanical systems, including incompressible and turbulent flows governed by the Navier--Stokes equations~\cite{Raissi2018_HFM_arXiv, jagtap2020extended, Jagtap2020_cPINN,Yonggana_POF2026} and also for solving Reynolds-averaged Navier--Stokes (RANS) equations, reconstructing turbulent fields from limited data, and incorporating experimental measurements into simulation frameworks~\cite{eivazi2022physics, eivazi2024physics, cai2021flow}. In recent years, microswimmers have been trained using adversarial reinforcement learning in forced two-dimensional (2D) Navier–Stokes turbulence~\cite{Alageshan_2020,
    gupta2025flockingaidpathplanning, gajendragad2026obstacleawarenavigationsmartmicroswimmers}.
 Conventionally constructed using multilayer perceptrons (MLPs), these networks integrate physical laws directly into the training process, enabling efficient solutions for forward and inverse problems.  The basic framework for scientific machine learning in fluid mechanics was established by Raissi et al., who introduced Physics-Informed Neural Networks (PINNs) to solve forward and inverse problems governed by nonlinear partial differential equations~\cite{Raissi2019_PINN}, and following this, expanded this approach to uncover hidden fluid mechanics from flow visualization data ~\cite{Raissi2018_HFM_arXiv}. To handle complex domains and improve parallelization, some domain decomposition strategies such as Conservative PINNs (cPINNs) ~\cite{Jagtap2020_cPINN} and Extended PINNs (XPINNs)~\cite{jagtap2020extended} were also developed. Kutz emphasised the transformative potential of deep learning in fluid dynamics, highlighting its capability to revolutionize turbulence modeling and flow control by bypassing the limitations of traditional closure schemes~\cite{kutz2017deep}. At the same time, Sirignano and Spilopoulos parallely introduced the Deep Galerkin Method (DGM), which is a meshfree algorithm that leverages deep neural networks to solve high-dimensional partial differential equations, addressing the curse of dimensionality inherent in conventional grid-based methods~\cite{sirignano2018dgm}. To improve the resolution of local features and sharp gradients which are often encountered in complex flow regimes—Kharazmi et al. developed hp-VPINNs, a variational framework incorporating domain decomposition and localized polynomial basis functions~\cite{kharazmi2021hp}. The theoretical basis of these approaches were rigorously examined by De Ryck and Mishra, whose numerical analysis of physics-informed machine learning models established critical error bounds and convergence guarantees, identifying the optimization challenges in stiff PDE problems~\cite{de2024numerical}. 

Even with these advances, conventional multilayer perceptrons used in PINNs pose certain problems. They often exhibit spectral bias \cite{Rahaman2019SpectralBias, Tancik2020FourierFeatures}, vanishing gradients \cite{pascanu2013difficulty}, and difficulties in capturing sharp gradients or memory effects in complex viscoelastic flows. To address these challenges, Kolmogorov-Arnold Networks (KANs) \cite{liu2024kan}, inspired by the Kolmogorov-Arnold representation theorem was proposed as an alternative to multi-layer perceptrons by Liu et al. These networks place learnable activation functions on edges rather than nodes, offering novel structural advantages and greater hyperparameter efficiency over traditional multi-layer perceptrons \cite{Cui_POF2025} due to their enhanced interpretability and improved accuracy in small-scale learning tasks. KANs also exhibit better interpretability and numerical conditioning, making them well suited for learning structured mappings in high-dimensional nonlinear systems.
The incorporation of physics-informed learning principles with KAN architectures (PI-KAN) creates a powerful hybrid framework for tackling viscoelastic fluid equations. This is where nonlinear velocity field evolution and relaxation mechanisms pose significant modeling challenges, which is solved by the PI-KAN.

LeCun et al. extensively reviewed the rapid advancement of deep learning  \cite{lecun2015deep}, which provided the computational basis for modern data-driven physics, thereby enabling the approximation of high-dimensional functions through multi-layer architectures \cite{Kovachki2023NeuralOperatorSurvey}. However, in the specific context of physics-informed learning, Wang et al. identified critical ``gradient flow pathologies". Wherein the numerical stiffness causes imbalances between the PDE residual loss and boundary condition gradients, which often lead to convergence failures in complex flow problems \cite{wang2021understanding, lu2021deepxde, Krishnapriyan2021FailureModes, Wang2022AdaptiveWeights, Yu2022GradBalance}. To address the modeling challenges inherent to rheologically complex materials, ``nn-PINNs," was introduceed by Mahmoudabadbozchelou et al. This is a framework tailored for non-Newtonian fluids that successfully captures shear-thinning and thickening behaviors using variationally consistent formulations ~\cite{mahmoudabadbozchelou2022nn}. Building further on this, for memory-dependent systems, Thakur et al. developed ``ViscoelasticNet," which is a specialized PINN framework designed to perform model selection for viscoelastic constitutive equations like the Oldroyd-B and FENE-P models by inferring the stress evolution in the system~\cite{thakur2024viscoelasticnet}. Despite these methodological improvements, traditional MLP-based backbones remain limited by spectral bias; very recent work by Khedr et al. has demonstrated that Physics-Informed Kolmogorov-Arnold Networks (PI-KANs) significantly outperform standard architectures in fluid simulations, offering superior accuracy and faster convergence by leveraging learnable activation functions on network edges~\cite{khedr2025physics, EvoKAN2025, PIKAN_JMLR2025}.

In our work, we develop a Physics-Informed Kolmogorov–Arnold Network (PI-KAN) framework for solving viscoelastic fluid equations, which involve nonlinear coupling between pressure, and velocity fields and are challenging for traditional numerical and machine learning approaches. To assess the performance of the proposed model, we consider benchmark problems based on the Taylor–Green (TG) flow and a modified Taylor–Green flow adapted for viscoelastic fluids~\cite{Colabrese_2017}.

We thoroughly investigate the influence of network architectures, hyperparameters, and collocation point distributions on the accuracy and convergence of PI-KANs for solving viscoelastic fluid equations. In addition, we analyze the behavior of multi-output KAN architectures when learning coupled physical variables and examine strategies to improve training stability. Numerical experiments demonstrate that the proposed PI-KAN framework accurately captures the dynamics of viscoelastic flows, achieving high prediction accuracy for velocity fields. We consider Generalised Hydrodynamic (GHD) model  to study viscoelastic fluid~\cite{kaw}.  Gupta et al. studied  viscoelastic memory effects on Kolmogorov flow in two-dimensional viscoelastic fluid using GHD model~\cite{aka6th, Aka4th}. 

The remainder of this paper is organized as follows: In
Sec.~\ref{Sec:model} we define our model and describe the numerical
methods we use. In Sec.~\ref{Sec:result} we present our results. Sec.~\ref{Sec:Conclusion} contains a discussion  of our results. 

\section{Models and Numerical methods}
\label{Sec:model}
\subsection{The Machine learning Model}

The development of data-driven and physics-informed approaches for solving complex nonlinear partial differential equations (PDEs) was primarily enabled by Scientific machine learning. Physics-Informed Neural Networks (PINNs) integrate governing equations, boundary conditions, and initial conditions directly into the loss function, allowing neural networks to learn physically consistent solutions with sparse data \cite{Raissi2019_PINN}. Fourier Neural Operator (FNO) and Deep Operator Network (DeepONet); which are Operator-learning frameworks, extend this paradigm beyond by learning mappings between function spaces, and enabling efficient solutions of parametric PDE families~\cite{li2020fourier, lu2021deeponet}. More recently, Kolmogorov--Arnold Networks (KANs) have emerged as a better alternative architecture, as they function by replacing node-based activations with learnable functional representations along edges. This improves the interpretability and approximation capabilities of the network \cite{liu2024kan}. These approaches have been promising fields like computational fluid dynamics, particularly for modeling turbulent flows and nonlinear transport phenomena.

Traditional numerical solvers for nonlinear PDEs are often computationally expensive, especially for high-dimensional and strongly nonlinear systems. In contrast, deep learning models can approximate solution operators and significantly accelerate computations. Architectures such as convolutional neural networks (CNNs)~\cite{krizhevsky2012imagenet,fukushima1980neocognitron,lecun1998gradient}  and recurrent neural networks (RNNs)~\cite{elman1990finding, hochreiter1997lstm, cho2014gru} have demonstrated the ability to capture complex spatial and temporal correlations~\cite{krizhevsky2012imagenet,fukushima1980neocognitron,lecun1998gradient,elman1990finding, hochreiter1997lstm, cho2014gru}.

Among neural architectures, multilayer perceptrons (MLPs) serve as fundamental building blocks and universal function approximators. An MLP consists of fully connected layers where each layer applies a linear transformation followed by a nonlinear activation:
\begin{equation}
\mathbf{a}^{(l)} = \sigma \left( \mathbf{W}^{(l)} \mathbf{a}^{(l-1)} + \mathbf{b}^{(l)} \right),
\end{equation}
where $\mathbf{a}^{(l-1)}$ represents the output of the previous layer, $\mathbf{W}^{(l)}$ and $\mathbf{b}^{(l)}$ are the weight matrix and the bias vector for layer l,and $\sigma(\cdot)$ denotes the activation function. $\mathbf{a}^{(l)}$ denotes the output of the current layer. MLPs can approximate complex nonlinear mappings but often require large datasets and may fail to satisfy physical laws when used purely in a data-driven manner. To address this limitation, PINNs embed physical constraints into the training process by constructing a composite loss function that includes PDE residuals and boundary/initial condition terms. This reduces the network's reliance on labeled data while at the same time ensuring physically consistent predictions. In the next subsections, we describe Kolmogorov Arnold Networks (KANs) and Physics Informed Kolmogorov Arnold Networks
(PI-KANs) briefly. 

\subsubsection{ Kolmogorov Arnold Networks (KANs)}
A much better methodology that has emerged in the recent days is that of KANs, which offer enhanced flexibility and interpretability~\cite{liu2024kan,Liu:PRX2025}. Inspired by the Kolmogorov Arnold Representation Theorem, these offer a unique design to model multivariate function with greater efficiency and accuracy. The theorem states that any continuous multivariate function 'f' defined on a bounded domain can be decomposed into a finite number of continuous univariate function and additive operations. The network constructed based on the Kolmogorov-Arnold representation theorem can be expressed as:\begin{equation}
f(\mathbf{x}) = \sum_{j=1}^{q} \phi_j \left( \sum_{i=1}^{n} g_{ij}(x_i) \right),
\end{equation}

 where $f(\mathbf{x})$ is the target multivariate function, $\mathbf{x} = (x_1, x_2, \dots, x_n)$ is the input vector, $q$ is the number of layers, $n$ is the number of input dimensions,
 $g_{ij}(x_i)$ are univariate transformation functions, and $\phi_j(\cdot)$ are nonlinear
 activation functions. The activation function was defined as: \begin{equation}
g(x_i) = b(x_i) + \sum_{k} \beta_k B_k(x_i),
\end{equation} a combination of a base function $b(x_i)$ and a spline function $B_k(x_i)$ in the original implementation. Later on learnable coefficients ($\beta_k$) were added to the equation to make it much more efficient, as shown above. The flexibility of this expression enable KANs to efficiently capture intricate nonlinearities.

\subsubsection{ Physics Informed Kolmogorov Arnold Networks (PI-KANs)}
  PI-KANs extend the capabilities of KANs by embedding physical laws, such as the one used here, the General Hydrodynamic Equations (GHD)~\cite{kaw,Gupta_POP2014} directly into the network's training process. This mixed approach combines the interpretability and flexibility of KANs with the rigorous enforcement of governing physical principles, making it a powerful framework for solving complex fluid dynamics problems. In a two-dimensional scenario, the PI-KAN model takes the spatial coordinates x and y as inputs and predicts the flow variables, including the velocity and vorticity components. Automatic differentiation is used to compute the derivatives of the network outputs, allowing the GHD equations to be formulated as residuals within the network's loss function. The PI-KAN model is designed to satisfy both the GHD equations and the prescribed boundary conditions, more of which is introduced later. These modern machine learning frameworks provide powerful tools for approximating solutions to nonlinear PDEs, with KANs offering a promising direction for improved accuracy and physical consistency in challenging applications.

\subsection{Architecture}

\begin{figure*}[t]  
    \centering
    \includegraphics[width=\textwidth]{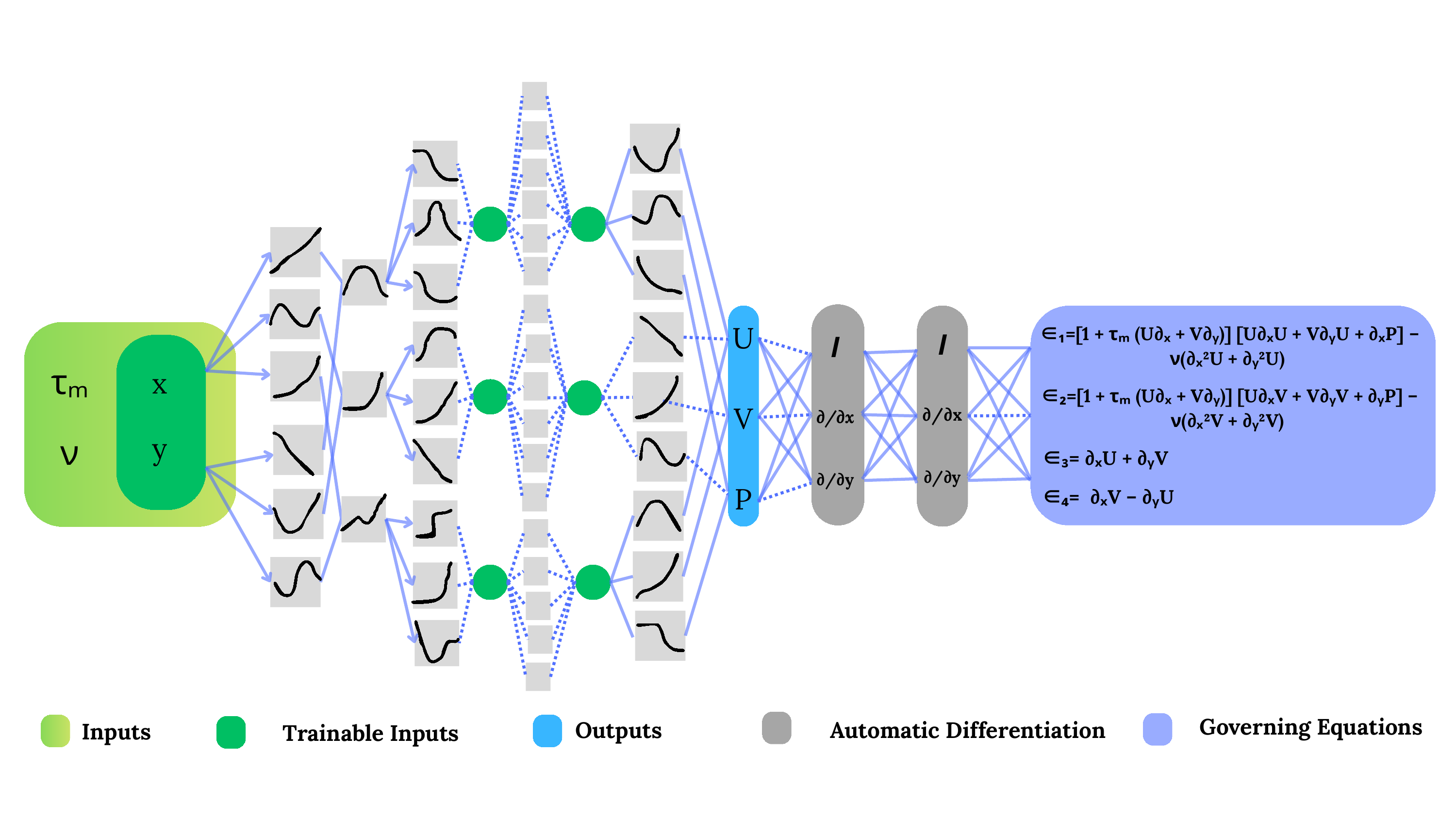}
    \caption{Schematic illustration of Physics-Informed Kolmogorov–Arnold Networks for solving viscoelastic fluid equations.}
    \label{fig:1}
\end{figure*}

The model basically uses Neural Networks but instead of the traditional MLP approach, we use KANs constrained by physics equations as discussed before. If an explicit solution to the PDE is available, the architecture of the KANs can be  designed by exploiting their interpretability advantages. The proposed PI-KAN leverages the Kolmogorov Arnold functional Decomposition theorem to represent multivariate mappings as compositions of learnable univariate functions. Unlike conventional MLPs that rely on fixed nonlinear activations, here each neuron in a KAN uses an adaptive B-spline basis expansion, allowing the activation itself to evolve during training. This setup enables the network to capture smooth, interpretable and higher order derivatives nonlinearities without explicit sparsification or pruning. To make sure it follows the Physics, the network is trained under a Physics Informed loss function comprising of the governing PDE residuals, boundary conditions, and periodicity constraints, computed using automatic differentiation. The model parameters are standardised using  $ L_1/L_2 $ normalization, where $L_1$ and $L_2$ normalization techniques are regularization methods used to prevent overfitting by adding a penalty term to a model's loss function. $L_1$ adds the sum of absolute weight values, while $L_2$ adds the sum of squared weights and optimization is performed using the Adaptive Movement Estimation (Adam) and limited memory BFGS (LBFGS) algorithms for improved convergence on tightly coupled residual landscapes.  This hybrid formulation ensures that the PI-KAN learns physically consistent solutions while maintaining interpretability through its spline based functional layers. 
We trained a large KAN with five layers structured as [[2,0], [5,5], [5,5], [5,5], [5,5], [2,0]], consisting of one input layer, four hidden layers, and one output layer. In addition to this, we uniformly distributed $41\times41$ points all over the computational domain. These points are required to enforce that the neural network satisfies the PDEs. These collocation points are used to impose the PDE constraints during training.
 The neural network settings are as follows: grid size is 3, and $k$ is 3. Here, the grid size denotes the number of spline intervals; that is the input range is divided into 3 intervals with its each unique polynomial behavior. '$k$' denotes the degree of the spline basis, here $k$ is 3, which means that the the $\beta$-splines that we are using in our neural network are of degree 3 or cubic. The KANs use splines along each edge to transform the data, where the splines are formed by linear combinations of basic $\beta$-splines. The grid parameter controls the number of $\beta$-splines, a larger grid value results in more number of B-splines, enabling finer control over composite $\beta$-splines.
This is represented schematically in Fig.~\ref{fig:1}. The inputs consist of all the variables that are needed to form and solve the equation. Out of these there are trainable parameters (Grid parameters; x and y coordinates), that are used to form the derivatives and second derivatives that are in the equation. Once the grid parameters are taken in, they are then fed to the Neural network, which here is in the form of a Kolmogorov Arnold Network (KAN). These are then fed into various layers which help in solving the physics equation given to the KAN network. The subsequent layers, with the help of $\beta$-splines, activation functions, and their linear combinations, are used to solve the equations. Each grid parameter is taken as a set $(x,y)$ and fed into the governing equations after which they are put up against the actual solutions, and then the loss or difference between the results calculated. Based on this, the model then backpropagates, to adjust the weights and biases, to minimize the loss. To resolve the distortions in vorticity due to high viscoelasticity, the network capacity is first expanded to a [2, 7, 7, 7, 7, 3] architecture to better capture the complex flow features.

\section{Fluid models}
\label{sec:fluid_model}
We consider viscoelastic fluid medium. The viscoelastic medium in the field of dusty plasma is commonly described using a generalized hydrodynamic (GHD) model~\cite{kaw,Gupta_POP2014, Aka4th}. This framework has previously been employed successfully by Kaw et al. to investigate transverse shear waves in strongly coupled dusty plasma~\cite{kaw}. Here, we present the governing equations for dust fluid dynamics, incorporating the effects of strong coupling. The evolution of dust density is described by the continuity equation.

\begin{equation}
\frac{\partial n}{\partial t} + \nabla \cdot \left( n \vec{v} \right) = 0,
\end{equation}

where $n$ is number density of the dust particle. The momentum equation has the generalized form to incorporate the viscoelastic effects

\begin{equation}
\begin{split}
\left[ 1 + \tau_m \left( \frac{\partial}{\partial t} + \vec{v} \cdot \nabla \right) \right]
\left[
\left( \frac{\partial}{\partial t} + \vec{v} \cdot \nabla \right)\vec{v} + \frac{\nabla P}{n} - \nabla \phi
\right]\\
= \nu \nabla^2 \vec{v},
\end{split}
\end{equation}

where $\vec{v}$, $P$, and $\phi$ are the dust velocity, pressure and potential respectively. The viscoelastic memory effect is incorporated through a relaxation time parameter $\tau_m$, and $\nu$ represents the viscosity coefficient. However, since dusty plasma is a charged medium, an additional coupling between the electrostatic potential and dust dynamics must be taken into account. In the incompressible flow limit, potential perturbations can be neglected. The two-dimensional incompressible viscoelastic fluid  equations with no body forces in the velocity-pressure formulation are expressed as in the incompressible limit:
\begin{equation}
\dfrac{\partial u}{\partial x} + \dfrac{\partial v}{\partial y} = 0,
\end{equation}
\begin{equation}
\begin{split}
\left(1+\tau_m \left(\dfrac{\partial }{\partial t}+ u \dfrac{\partial}{\partial x} + v \dfrac{\partial }{\partial y}\right)\right) \left(\dfrac{\partial u}{\partial t} + u \dfrac{\partial u}{\partial x} + v \dfrac{\partial u}{\partial y}+\dfrac{\partial P}{\partial x}\right)\\
= \nu \left( \dfrac{\partial^2 u}{\partial x^2} + \dfrac{\partial^2 u}{\partial y^2} \right),
\end{split}
\end{equation}
\begin{equation}
\begin{split}
\left(1+\tau_m (\dfrac{\partial }{\partial t}+ u \dfrac{\partial}{\partial x} + v \dfrac{\partial }{\partial y})\right) \left(\dfrac{\partial v}{\partial t} + u \dfrac{\partial v}{\partial x} + v \dfrac{\partial v}{\partial y}+\dfrac{\partial P}{\partial y}
\right)\\
= \nu \left( \dfrac{\partial^2 v}{\partial x^2} + \dfrac{\partial^2 v}{\partial y^2} \right),
\end{split}
\label{eq:NS}
\end{equation}
where $u$ and $v$ are the velocity components in the $x$ and $y$ directions, respectively. 
\medskip
The canonical Taylor-Green vortex serves as a fundamental benchmark for evaluating dissipative mechanisms and turbulent transitions in continuous and neural-network-based fluid models \cite{TaylorGreen1937, jin2021nsfnets, wang2025fluctuating}, whose underlying rheological behaviors are rigorously defined in classical viscoelastic frameworks \cite{Bird1987DynamicsPolymeric, Larson1999Constitutive}. For the Taylor-Green vortex, the velocity components are defined as
\begin{equation}
\begin{cases}
u(x, y, t) = \sin(x)\cos(y)e^{-2\nu t}, \\[6pt]
v(x, y, t) = -\cos(x)\sin(y)e^{-2\nu t},
\end{cases}
\label{eq:TGV}
\end{equation}
These analytical solutions satisfy both the continuity and momentum equations, providing an exact representation of the flow field in the Navier-Stokes limit ($\tau_m=0$). It is important to note that the Taylor–Green vortex does not constitute a solution to the viscoelastic fluid equation. Therefore, the evaluation of the PDE loss is carried out using an alternative formulation. Further details regarding the PDE loss calculation are provided in Sec.~\ref{sec:pikans}.
In this study, we focus on a two-dimensional implementation of the Taylor--Green vortex for validation purposes. 
The computational domain is defined as $[0, 2\pi] \times [0, 2\pi]$, with $x$ and $y$ representing the Cartesian coordinates. 
For simplicity, the fluid's temporal behavior is fixed at $t=0$, yielding the initial velocity components
\begin{equation}
u(x, y) = \sin(x)\cos(y), \qquad v(x, y) = -\cos(x)\sin(y).
\label{eq:initial}
\end{equation}

To probe the robustness of the PI-KAN framework beyond the canonical Taylor--Green configuration, we consider a class of composite velocity fields designed to introduce controlled multi-scale perturbations and break the inherent spatial symmetry of the baseline flow. Let $(u_1, v_1)$ denote the standard Taylor--Green velocity field defined in Eq.~(12). A secondary flow component $(u_2, v_2)$, characterized by higher spatial frequencies and a phase shift, is introduced as
\begin{equation}
\begin{cases}
u_2(x,y) = \sin\!\left(2(x-\Delta x)\right)\cos\!\left(2(y-\Delta y)\right),\\[4pt]
v_2(x,y) = -\cos\!\left(2(x-\Delta x)\right)\sin\!\left(2(y-\Delta y)\right),
\end{cases}
\label{eq:perturbed_velocity}
\end{equation}
where the phase offsets are fixed at $(\Delta x, \Delta y) = (3.35,\,1.83)$. The total velocity field is constructed through a linear interpolation controlled by a mixing parameter $\beta \in [0,1]$,
\begin{equation}
u = \beta u_1 + (1-\beta)u_2, \qquad
v = \beta v_1 + (1-\beta)v_2.
\label{eq:velocity_mixture}
\end{equation}
This parametrization enables a smooth transition from the canonical Taylor--Green flow ($\beta = 1$) to increasingly distorted configurations as $\beta$ decreases. The corresponding vorticity field, defined as $\omega = \partial v/\partial x - \partial u/\partial y$, follows the same linear structure,
\begin{equation}
\omega(x,y) = \beta \omega_1 + (1-\beta)\omega_2,
\label{eq:vorticity_mixture}
\end{equation}
where $\omega_1 = 2\sin(x)\sin(y)$ is the vorticity associated with the baseline Taylor--Green vortex and $\omega_2 = 4\sin\!\left(2(x-\Delta x)\right)\sin\!\left(2(y-\Delta y)\right)$ corresponds to the perturbed mode. This specific formulation provides a systematic mechanism to introduce controlled structural complexity in the flow, without jeopardizing analytical tractability, and allowing a clear assessment of the model's performance under progressively deformed flow conditions.

\section{PI-KANs for viscoelastic fluid equations}
\label{sec:pikans}
The PI-KAN model is designed to satisfy both the viscoelastic fluid equations and the
prescribed boundary conditions. The total loss function is defined
as
\begin{equation}
\mathcal{L} = \mathcal{L}_{\text{PDE}_1}+\mathcal{L}_{\text{PDE}_2} +\mathcal{L}_{\text{PDE}_3}  + \mathcal{L}_{\text{BC}}, 
\end{equation}
where $\mathcal{L}_{\text{PDE}}$ quantifies the residuals of the fluid equations across the computational domain, and $\mathcal{L}_{\text{BC}}$ measures the error in satisfying boundary conditions. The PDE residual loss ($\mathcal{L}_{\text{PDE}}$) is defined as
\begin{equation}
\mathcal{L}_{\text{PDE}_1} = 
\frac{1}{N_{\text{col}}}
\sum_{i=1}^{N_{\text{col}}}
\mathcal{R}(\mathbf{x}_i)^2 ,
\end{equation}

\begin{equation}
\label{eq:pde2}
\mathcal{L}_{\text{PDE}_2} = 
\frac{1}{N_{\text{col}}}
\sum_{i=1}^{N_{\text{col}}}
\left\| {R'}_i - {R'}_{i,\text{true}} \right\|^2, 
\end{equation}

\begin{equation}
\label{eq:pde3}
\mathcal{L}_{\text{PDE}_3} = 
\frac{1}{N_{\text{col}}}
\sum_{i=1}^{N_{\text{col}}}
\left\| {R''}_i - {R''}_{i,\text{true}} \right\|^2,
\end{equation}
where $\mathcal{R}$, $\mathcal{R'}$, and $\mathcal{R''}$ are the residuals of the continuity and momentum equations at $N_{col}$ collocation points within the domain.  $\mathcal{R'}_i$ and $\mathcal{R''}_i$ represents the predicted residual values, and 
$\mathcal{R'}_{i,\text{true}}$ and $\mathcal{R''}_{i,\text{true}}$ denote the known residual values at $N_{\text{col}}$ collocation points within the domain. The expressions to calculate $\mathcal{R'}_{i,\text{true}}$ and $\mathcal{R''}_{i,\text{true}}$ are given below. 

\begin{equation}
\begin{split}
\mathcal{R'}_{i,\text{true}}=\left(1+\tau_m \left( u_{i} \dfrac{\partial}{\partial x_{i}} + v_{i} \dfrac{\partial }{\partial y_{i}}\right)\right) \times \\\left(u_{i} \dfrac{\partial u_{i}}{\partial x_{i}} + v \dfrac{\partial u_{i}}{\partial y_{i}}+\dfrac{\partial P_{i}}{\partial x_{i}}\right)-\nu \left( \dfrac{\partial^2 u_{i}}{\partial x_{i}^2} + \dfrac{\partial^2 u_{i}}{\partial y_{i}^2} \right),
\end{split}
\end{equation}

\begin{equation}
\begin{split}
\mathcal{R''}_{i,\text{true}}=\left(1+\tau_m ( u_i \dfrac{\partial}{\partial x_i} + v \dfrac{\partial }{\partial y})\right) \times \\ \left( u \dfrac{\partial v_i}{\partial x_i} + v_i \dfrac{\partial v_i}{\partial y_i}+\dfrac{\partial P_i}{\partial y_i}
\right)-\nu \left( \dfrac{\partial^2 v_i}{\partial x_i^2} + \dfrac{\partial^2 v_i}{\partial y_i^2} \right)
\end{split}
\label{eq:Resudal2}
\end{equation}

We  calculate  $\mathcal{L}_{\text{PDE}_2}$ and $\mathcal{L}_{\text{PDE}_3}$ in Eq.~\ref{eq:pde2} and Eq.~\ref{eq:pde3} as Taylor-Green flow and modified Taylor-Green flow are not the solution of viscoelastic fluid equation. We describe the flows and viscoelastic fluid equations in the next section Sec.~\ref{sec:fluid_model}.
\begin{equation}
\mathcal{L}_{\text{BC}} = 
\frac{1}{N_{\text{BC}}}
\sum_{j=1}^{N_{\text{BC}}}
\left\| \mathbf{u}_j - \mathbf{u}_{j,\text{true}} \right\|^2,
\end{equation}

where $\mathbf{u}_j$ represents the predicted boundary values, and 
$\mathbf{u}_{j,\text{true}}$ denotes the known boundary conditions 
at $N_{\text{BC}}$ boundary points. To consolidate the momentum residuals, we define the combined momentum partial differential equation (MPDE) loss as:
\begin{equation}
\label{eq:mpde}
\mathcal{L}_{\text{MPDE}} = \mathcal{L}_{\text{PDE}_2} + \mathcal{L}_{\text{PDE}_3}.
\end{equation}
Furthermore, for the continuity equation, we denote its residual loss as the divergence loss ($\mathcal{L}_{\text{div}} \equiv \mathcal{L}_{\text{PDE}_1}$). During the training of physics-informed architectures, the gradients of different loss components can vary by several orders of magnitude, often leading to stiff optimization landscapes. To mitigate this and balance the contributions of each physical constraint, we introduce dynamic weighting parameters to the total loss function:
\begin{equation}
\label{eq:weighted_total_loss}
\mathcal{L} = \lambda_{\text{div}}\mathcal{L}_{\text{div}} + \lambda_{\text{MPDE}}\mathcal{L}_{\text{MPDE}} + \lambda_{\text{BC}}\mathcal{L}_{\text{BC}} + \lambda_{\omega}\mathcal{L}_{\omega},
\end{equation}
where $\lambda_{\text{div}}$, $\lambda_{\text{MPDE}}$, and $\lambda_{\text{BC}}$ are the weighting parameters corresponding to the divergence, momentum, and boundary condition losses, respectively. Additionally, $\lambda_{\omega}$ is introduced as a weighting parameter for an auxiliary vorticity loss ($\mathcal{L}_{\omega}$), which acts as a physical regularizer to ensure the network accurately captures the rotational dynamics and small-scale eddy structures inherent to the viscoelastic Taylor-Green flow.

\begin{table}[htbp]
\centering
\renewcommand{\arraystretch}{1.2} 
\begin{tabular}{l c l}
\hline\hline
\textbf{Symbol} & \textbf{Value} & \textbf{Description} \\
\hline
$\nu$           & $0.01$         & kinematic viscosity \\
$L$             & $2\pi$         & system size / domain length \\
$dx_{shift}$    & $3.35$         & spatial shift in x-axis \\
$dy_{shift}$    & $1.83$         & spatial shift in y-axis \\
$N_{train}$     & $41$           & training grid resolution ($41^2$) \\
$N_{test}$      & $256, 512$     & testing grid resolutions \\
$N_{p}$         & $84$           & periodic boundary sampled points \\
$G$             & $3$            & MultKAN grid intervals \\
$k$             & $3$            & MultKAN cubic B-spline order \\
Seed            & $32$           & random initialization seed \\ 
\hline\hline
\end{tabular}
\caption{List of constants and training parameter values.}
\label{tab:parameters}
\end{table}
 In the next section, we will discuss about our obtained results is detail. 

\section{Results}
\label{Sec:result}

\begin{figure*}[!htbp]  
    \centering
    \includegraphics[width=1.0\textwidth]{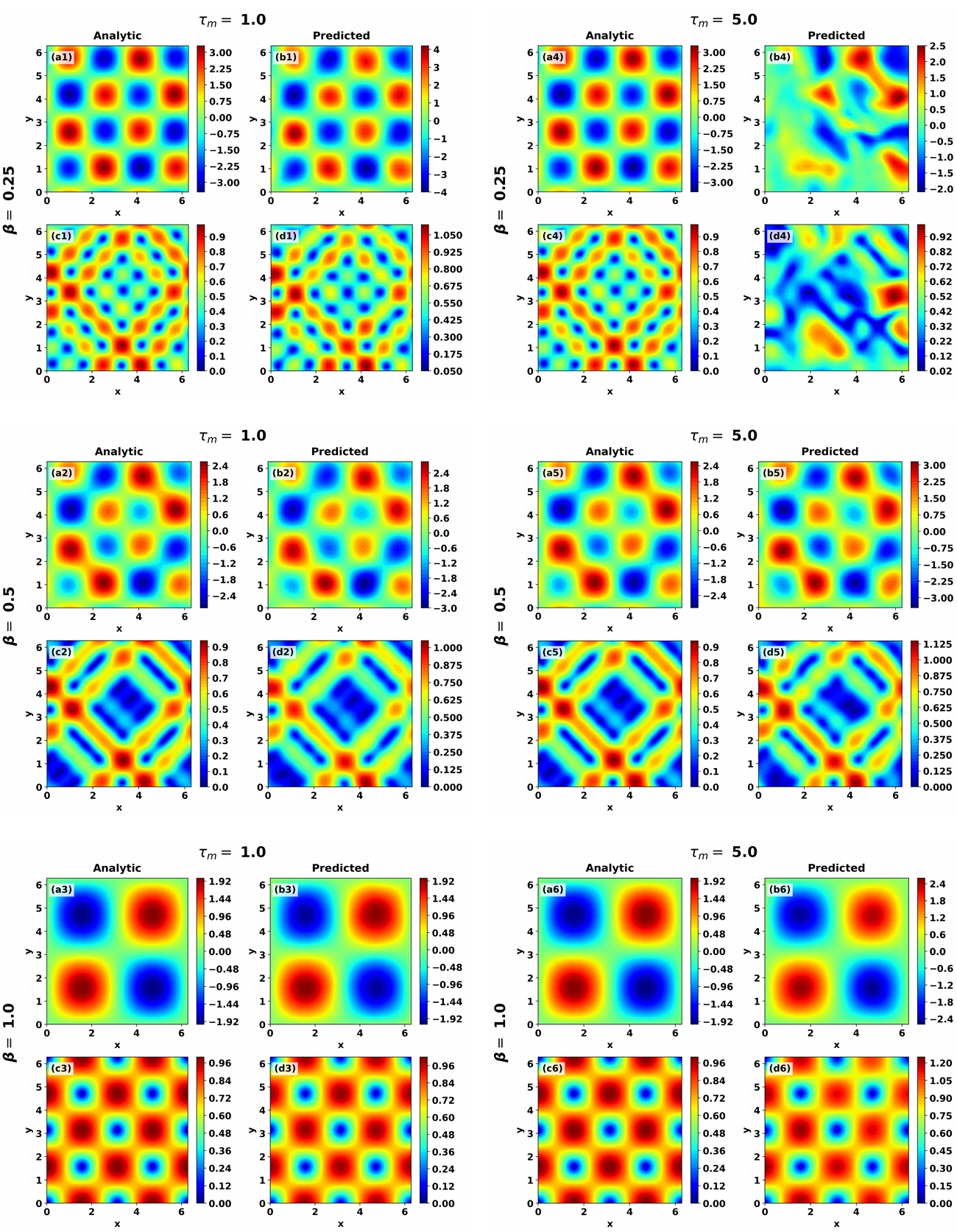}
    \caption{Comparison of the analytical and predicted flow fields across various viscoelastic relaxation times ($\tau_m$) and mixing parameters ($\beta$). Panels (a1-a6) and (c1-c6) display the analytical vorticity and velocity fields, respectively. Panels (b1-b6) and (d1-d6) show the corresponding MultKAN predictions. MultKAN architecture [2, 5, 5, 5, 5, 3] trained on a $41\times41$ grid, using loss weights $\lambda_{\text{BC}}=10$, and $\lambda_{\omega}=1$, initialized with Adam (300 steps, $\eta=10^{-3}$) and fine-tuned with LBFGS (20 restarts, 20 steps/restart, $\eta=1.0$ ).}
    \label{fig:2}
\end{figure*}
 
\begin{figure}[!htbp]
   \centering
   \includegraphics[width=1.0\linewidth]{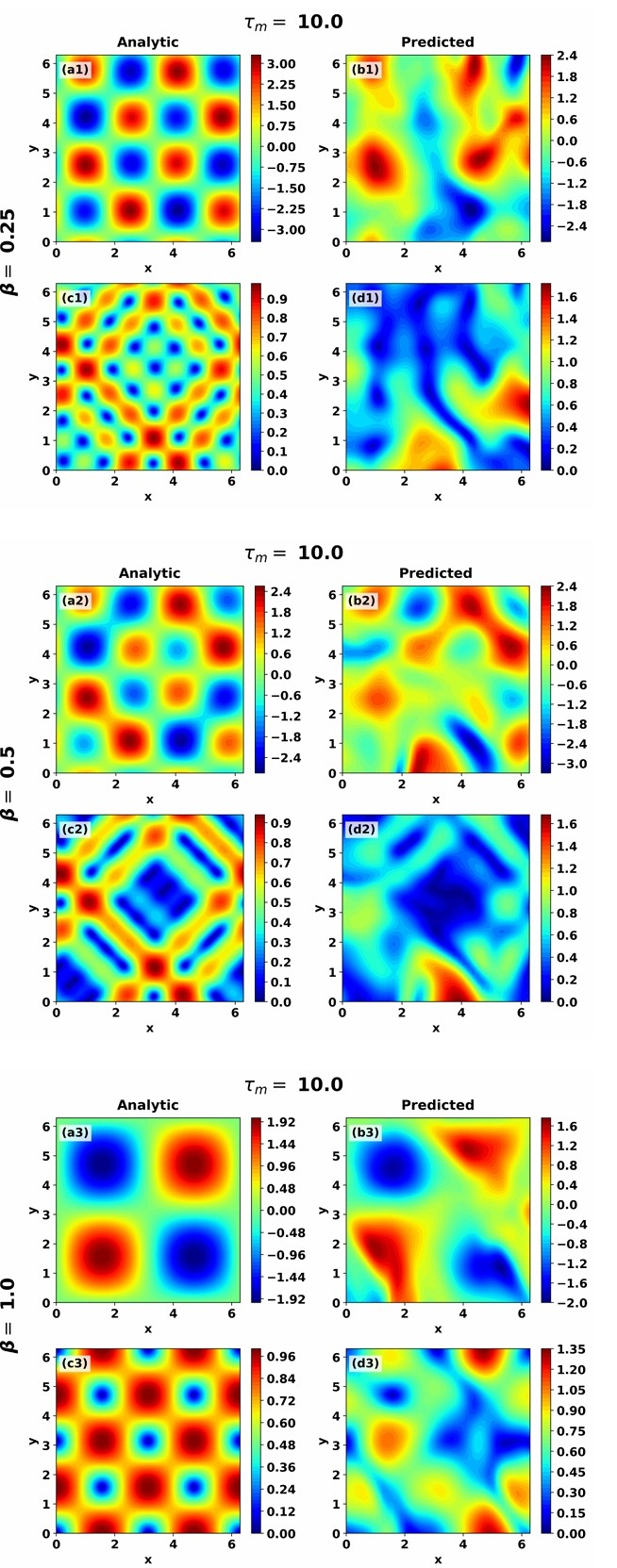}
   \caption{Comparison of the analytical and predicted flow fields across various viscoelastic relaxation times ($\tau_m$) and mixing parameters ($\beta$). Panels (a1-a3) and (c1-c3) display the analytical vorticity and velocity fields, respectively. Panels (b1-b3) and (d1-d3) show the corresponding MultKAN predictions}
   \label{fig:3}
\end{figure}

\begin{figure}[!htbp]
\centering
\includegraphics[width=0.5\textwidth]{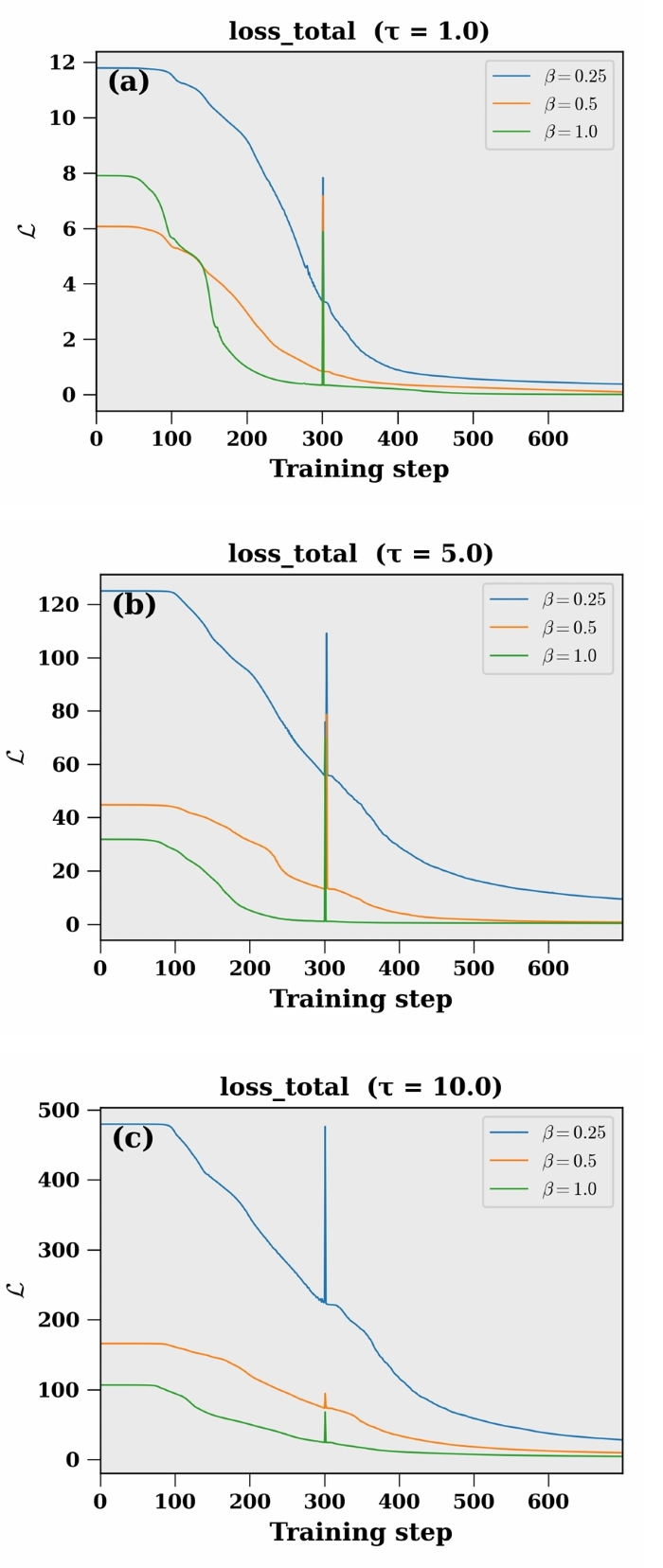}
\caption{Total loss vs training steps plot for $\tau_m=1 , 5, 10$ before gradient clipping with MultKAN architecture [2, 5, 5, 5, 5, 3]. }
\label{fig:loss_before_tau1_tau5_tau10}
\end{figure}

 The list of constants and training parameter for our simulation are given in Table.~\ref{tab:parameters}.
 Fig.~\ref{fig:2} represents a comparison between the analytical and predicted flow fields for different values of the visco elastic parameter $\tau_m$ and mixing parameter $\beta$.Within each configuration panel, the top row displays the analytical (a) and predicted (b) vorticity fields, while the bottom row illustrates the corresponding analytical (c) and predicted (d) velocity fields. For low viscosity ($\tau_m$ = 1.0), the PI-KAN model accurately reconstructs the flow structure across all $\beta$ values, with excellent agreement in both spatial patterns and magnitudes. In this figure, the MultKAN architecture [2, 5, 5, 5, 5, 3] trained on a $41 \times 41$ grid, using loss weights $\lambda_{BC}=10$, and $\lambda_{\omega}=1$.
The optimization was initialized with the Adam optimizer (300 steps, $\eta = 10^{-3}$) and fine-tuned with L-BFGS (20 restarts, 20 steps/restart, $\eta = 1.0$), where $\eta$ denotes the learning rate controlling the optimization step size. However for higher viscoelasticity, as $\tau_m=5.0$, noticeable deviations emerge, particularly for lower $\beta$ values ($\beta$ = 0.25, 0.5), where the flow consists of multiscale features. As shown in Fig.~\ref{fig:3}, the machine learning model fails to accurately predict both the velocity and vorticity fields.  In these cases, the predicted fields exhibit mild distortion and reduced symmetry, indicating difficulty in capturing strongly nonlinear viscoelastic effects. The model maintains better agreement for $\beta$ = 1.0 even at higher $\tau_m$. The vorticity structures remain relatively well preserved compared to velocity fields, indicating that the rotational features are more robustly captured by the network. In Fig.~\ref{fig:loss_before_tau1_tau5_tau10} , the total loss is plotted for $\tau_ m=1$, $\tau_ m=5$  and $\tau_m=10$. It is clearly demonstrated that a peak appears when L-BFGS begins immediately after the Adam training phase. The plot has been smoothed using gradient clipping.  

\begin{figure}[!htbp]
\centering
\includegraphics[width=0.5\textwidth]{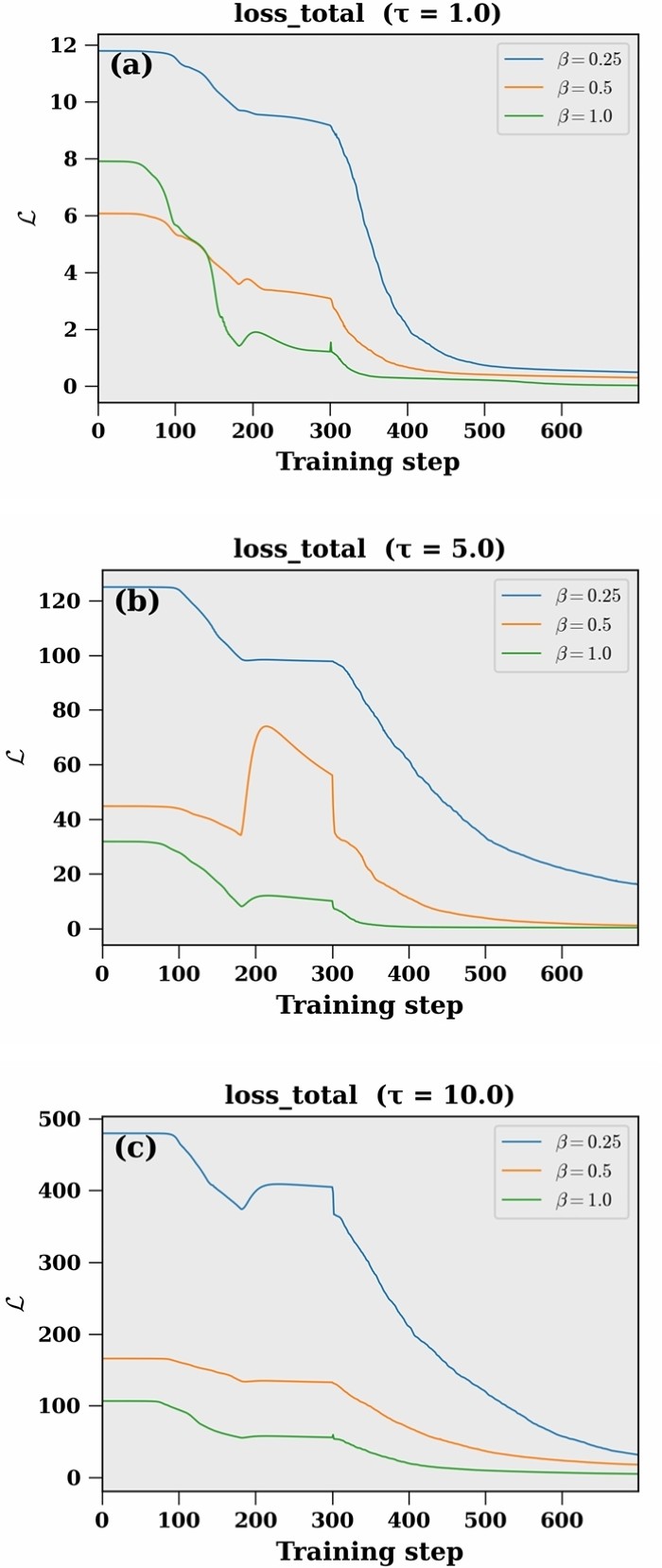}
\caption{Total Loss with gradient clipping for $\tau_m=1 , 5, 10$ with MultKAN architecture [2, 5, 5, 5, 5, 3].}
\label{fig:loss_with_GC}
\end{figure}

\begin{figure*}[!htbp]
\centering
\includegraphics[width=\linewidth]{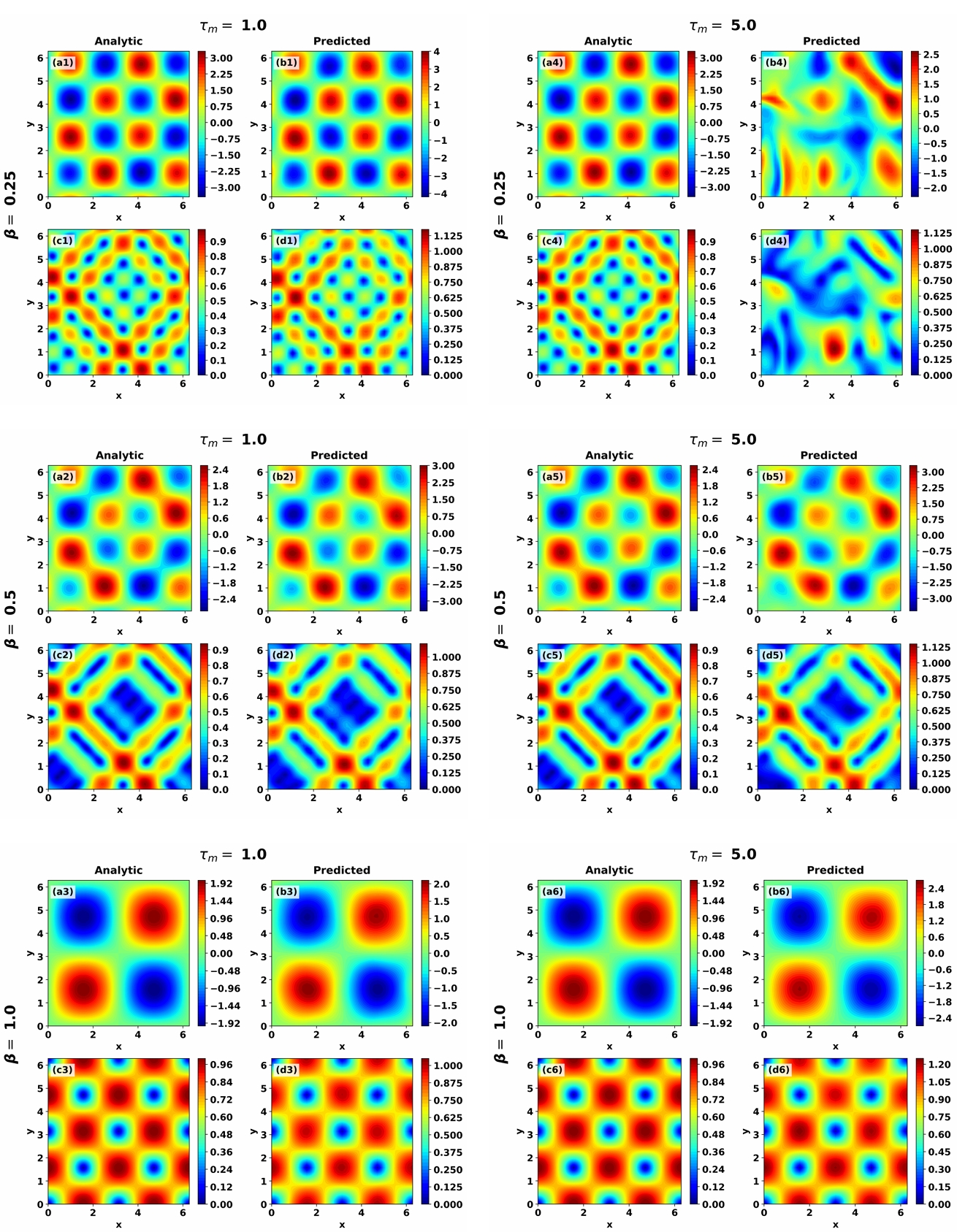}

\caption{Comparison of the analytical and predicted flow fields across various viscoelastic relaxation times ($\tau_m$) and mixing parameters ($\beta$). Panels (a1-a6) and (c1-c6) display the analytical vorticity and velocity fields, respectively. Panels (b1-b6) and (d1-d6) show the corresponding MultKAN predictions, stabilized using gradient clipping.
MultKAN architecture [2, 5, 5, 5, 5, 3] trained on a $41\times41$ grid, using loss weights $\lambda_{\text{BC}}=10$, and $\lambda_{\omega}=1$, initialized with Adam (300 steps, $\eta=10^{-3}$) and fine-tuned with LBFGS (20 restarts, 20 steps/restart, $\eta=1.0$ ). Gradient Clipping max norm = $0.05$ }
\label{fig:vorticity_with_GC}
\end{figure*}

Gradient clipping serves as an essential numerical stabilization technique by strictly capping the gradient norm at a predefined threshold \cite{bengio2017deep}. During the training of complex models, particularly within highly non-linear optimization landscapes, the presence of steep error ``walls'' can frequently trigger exploding gradients. The presence of such extreme values causes the optimizer to make excessively large weight updates, ultimately compromising the stability of the learning process. By restricting the maximum update size, clipping effectively confines the optimizer to smoother, low-curvature regions of the loss surface \cite{pascanu2013difficulty}. Gradient clipping plays an important role in stabilizing the training of Physics-Informed Neural Networks (PINNs), particularly for stiff governing equations where large gradients can hinder convergence in complex fluid simulations \cite{wang2021understanding}. By limiting excessively large updates during optimization, the training process becomes more stable and consistent. Fig.~\ref{fig:loss_with_GC} presents the total loss obtained after applying gradient clipping. The corresponding analytical and predicted vorticity and velocity fields are shown in Fig.~\ref{fig:vorticity_with_GC} for different values of $\beta$. In the following analysis, gradient clipping is employed throughout all simulations.

 To resolve the distortions at the viscoelastic parameter $\tau_m = 5.0$ and $\tau_m = 10.0$, the network architecture has been investigated to [2, 7, 7, 7, 7, 3] to better capture the complex flow features. To handle the numerical stiffness of the viscoelastic equations, we adopted a continuation strategy using dynamic PDE loss weighting. The momentum PDE weight was defined as $\lambda_{MPDE} = 0.1\alpha$, where $\alpha$ increases linearly from 0 to 1 between training steps 500 and 2500, providing a warmup phase during training. By keeping the PDE weight small in the initial stage, the network first learns the main flow structures under strongly weighted boundary ($\lambda_{BC} = 100$) and vorticity ($\lambda_{\omega} = 20$) constraints. After the initial flow structure is learned, the viscoelastic PDE constraints are gradually increased and fully enforced.

\begin{figure}[!htbp]
\centering
\includegraphics[width=1.0\linewidth]{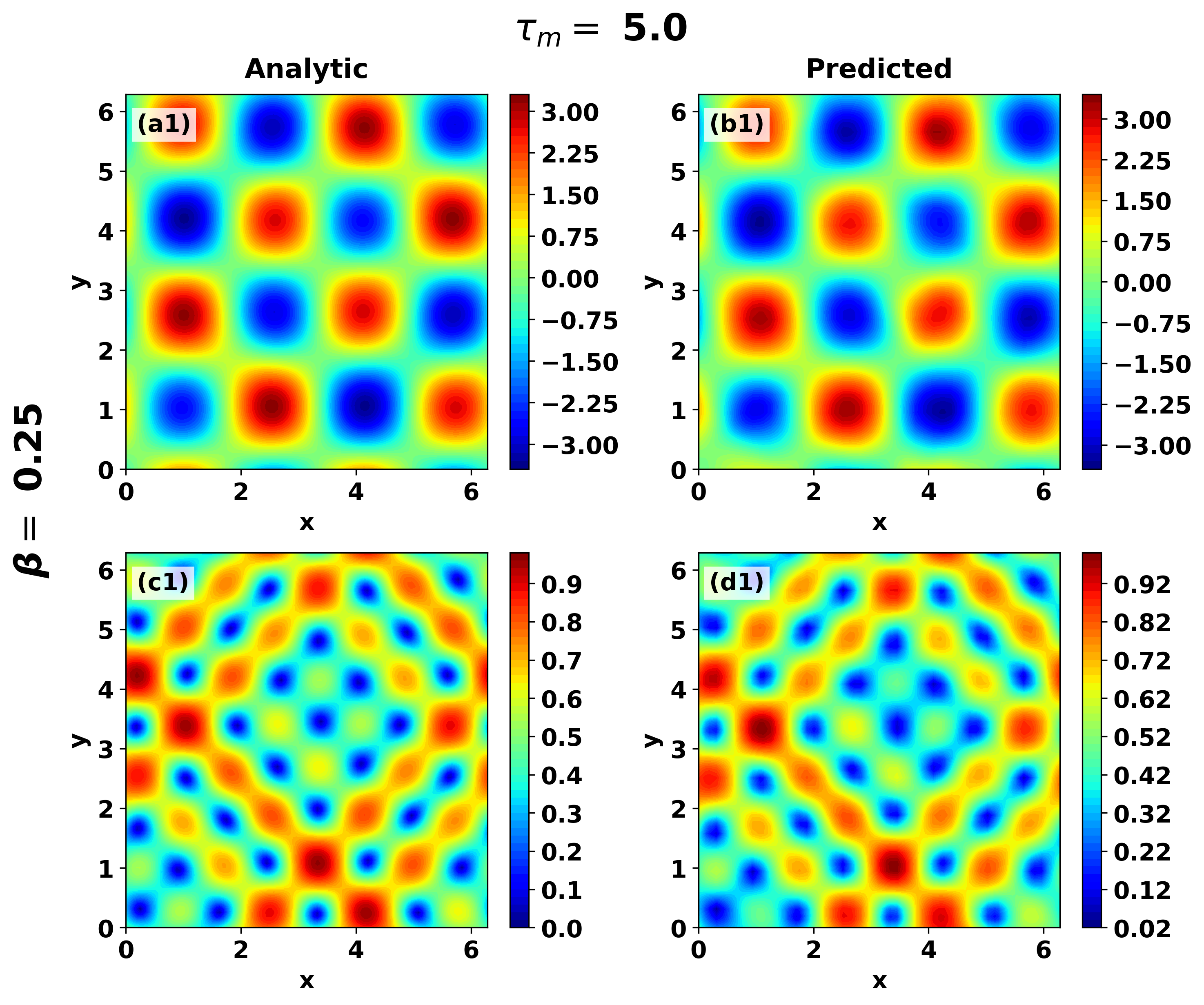}
\caption{Comparison of the analytical and predicted flow fields across various viscoelastic relaxation time ($\tau_m=5$) and mixing parameters ($\beta=0.25$). Panels a1 and c1 display the analytical vorticity and velocity fields, respectively. Panels b1 and d1 show the corresponding MultKAN predictions with architecture [2, 7, 7, 7, 7, 3], stabilized using gradient clipping }
\label{fig:6}
\end{figure}

\begin{figure}[!htbp]
\centering
\includegraphics[width=1.0\linewidth]{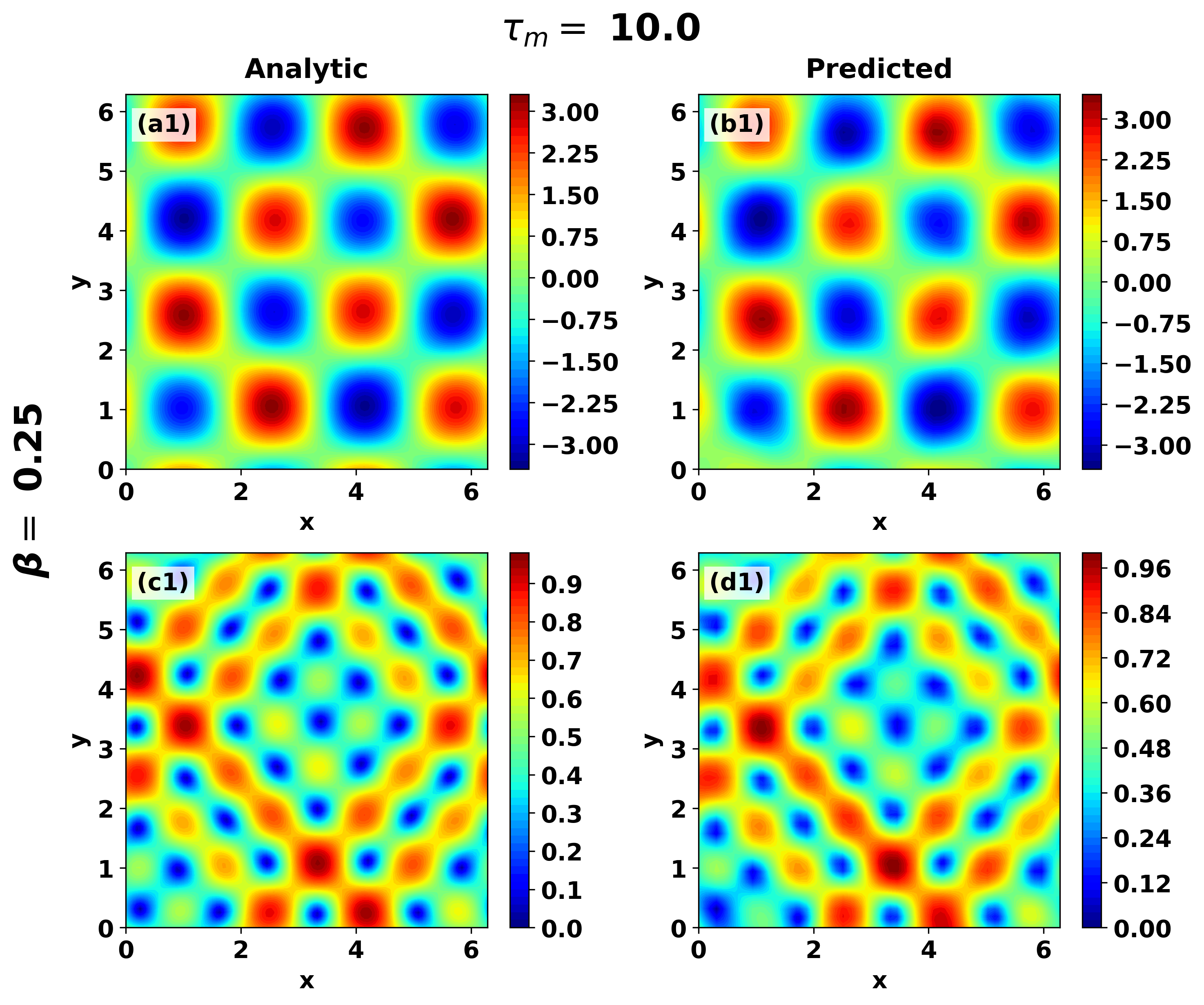}
\caption{Comparison of the analytical and predicted flow fields across various viscoelastic relaxation time ($\tau_m=10$) and mixing parameters ($\beta=0.25$). Panels a1 and c1 display the analytical vorticity and velocity fields, respectively. Panels b1 and d1 show the corresponding MultKAN predictions with architecture [2, 7, 7, 7, 7, 3], stabilized using gradient clipping }
\label{fig:7}
\end{figure}

Fig.~\ref{fig:6} and Fig.~\ref{fig:7}, show that the predicted vorticity and velocity fields for $\tau_m = 5.0$ and $\tau_m = 10.0$ agree well with the analytical solutions for $\beta = 0.25$. Vorticity and velocity fields  with $\beta = 0.25$ corresponds to the steepest gradients in the present setup. Stable predictions in this case indicate that the remaining $\beta$ configurations have already been resolved.

\begin{figure}[!htbp]
\centering
\includegraphics[width=1.0\linewidth]{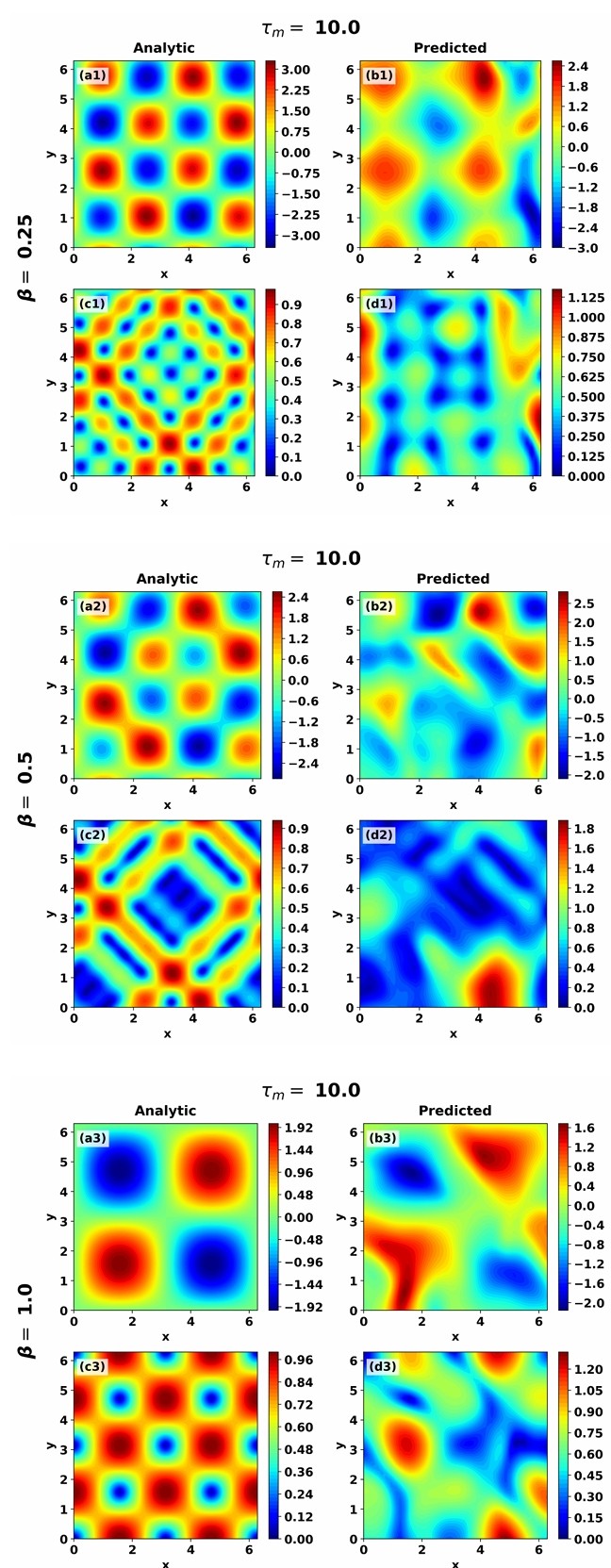}
\caption{{Comparison of the analytical and predicted flow fields across various viscoelastic relaxation times ($\tau_m$) and mixing parameters ($\beta$). Panels (a1-a3) and (c1-c3) display the analytical vorticity and velocity fields, respectively. Panels (b1-b3) and (d1-d3) show the corresponding MultKAN predictions, stabilized using gradient clipping.}}
\label{fig:vorticity_tua10_with_GC}
\end{figure}

\begin{figure}[!htbp]
\centering
\includegraphics[width=1.0\linewidth]{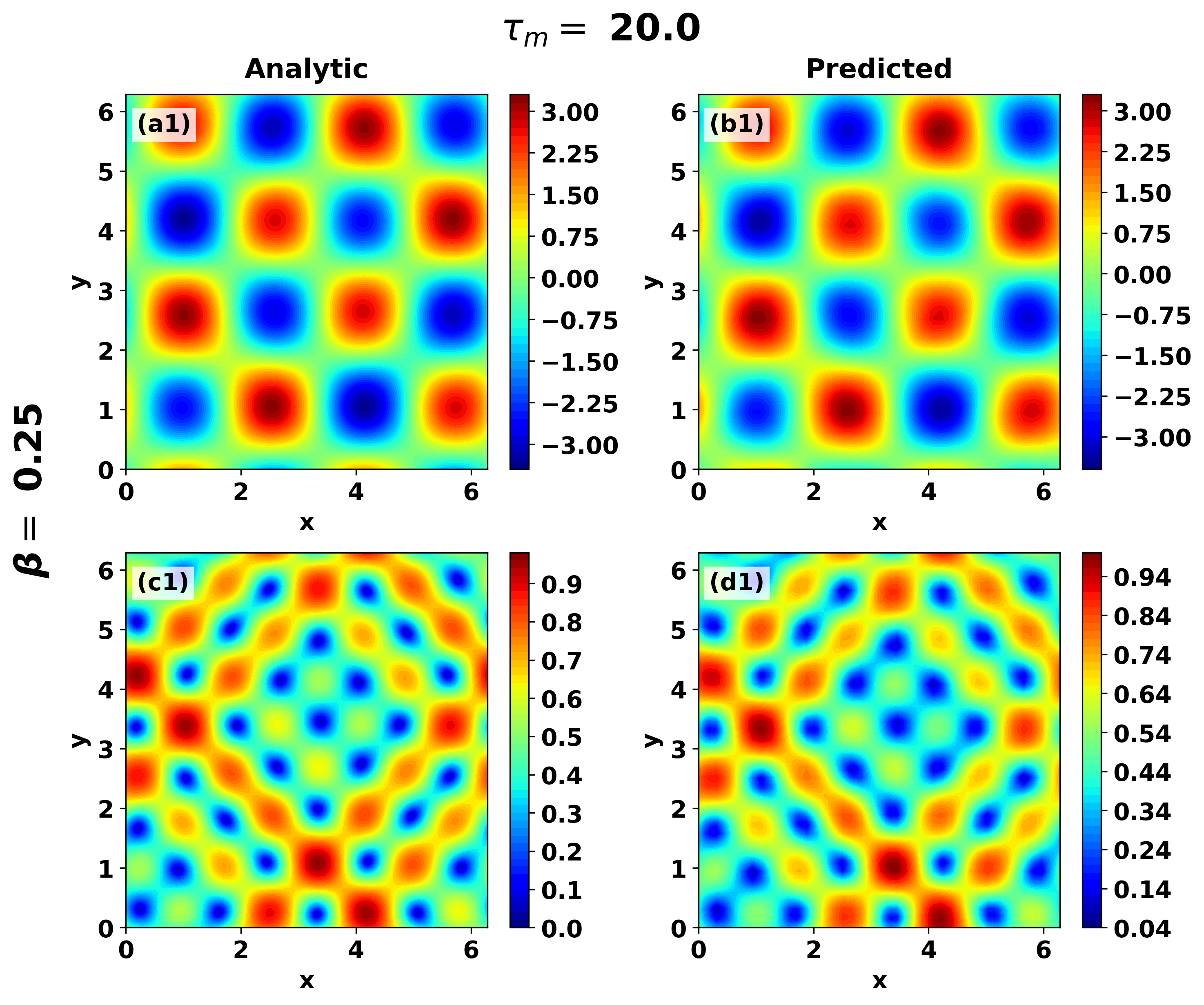}
\caption{Comparison of the analytical and predicted flow fields across various viscoelastic relaxation time ($\tau_m=20$) and mixing parameters ($\beta=0.25$). Panels a1 and c1 display the analytical vorticity and velocity fields, respectively. Panels b1 and d1 show the corresponding MultKAN predictions with architecture [2,  9, 9, 9, 9, 3], stabilized using gradient clipping }
\label{fig:tau=20}
\end{figure}

Pushing the high viscoelasticity further to a highly stiff regime of $\tau_m = 20.0$ reintroduced wobbly contours and aliasing when using the previously resolved configuration. To counteract this, the network was further widened to a [2, 9, 9, 9, 9, 3] configuration to provide the degrees of freedom necessary to map the sharper stress gradients. Crucially, the maximum dynamic PDE weight was scaled down to $\lambda_{\text{MPDE}} = 0.02\alpha$. This strict weight reduction balances the gradients during backpropagation, preventing the exploding viscoelastic residuals from destroying the spatial accuracy of the solution. As demonstrated in Fig. \ref{fig:tau=20}, these targeted architectural and hyperparameter adjustments successfully resolved the flow field for $\tau_m = 20.0$ and $\beta = 0.25$ without any visible distortion. 

\begin{figure}[!htbp]
\centering
\includegraphics[width=1.0\linewidth]{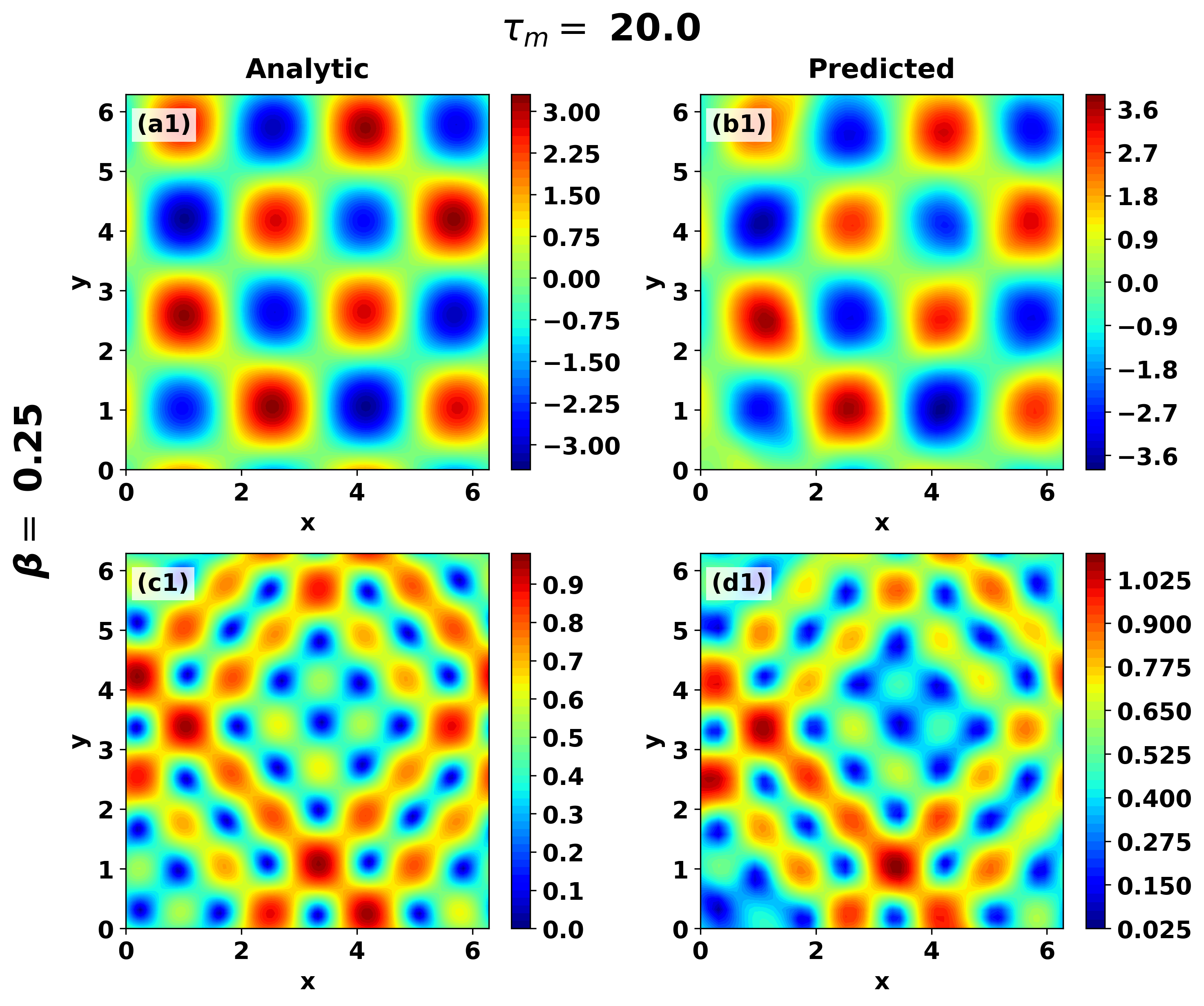}
\caption{Comparison of the analytical and predicted flow fields demonstrating the model's robustness to variations in kinematic viscosity ($\nu = 0.1$) within the high-viscoelasticity regime ($\tau_m = 20.0$, $\beta = 0.25$). Panels (a1) and (c1) display the analytical vorticity and velocity fields, respectively. Panels (b1) and (d1) show the corresponding MultKAN predictions. }
\label{fig:nu=0.1}
\end{figure}

 Finally, the robustness of the converged $\tau_m = 20.0$ model was evaluated by altering the kinematic viscosity in Fig.~\ref{fig:nu=0.1}. Simulations were run comparing $\nu = 0.01$ against an increased viscosity of $\nu = 0.1$. The predicted flow fields and spatial contours remained identical between the two cases, demonstrating that the network's predictive stability in this high-viscoelasticity regime is robust and independent of these specific variations in kinematic viscosity.

\begin{figure*}[!htbp]  
    \centering
    \includegraphics[width=1.0\textwidth]{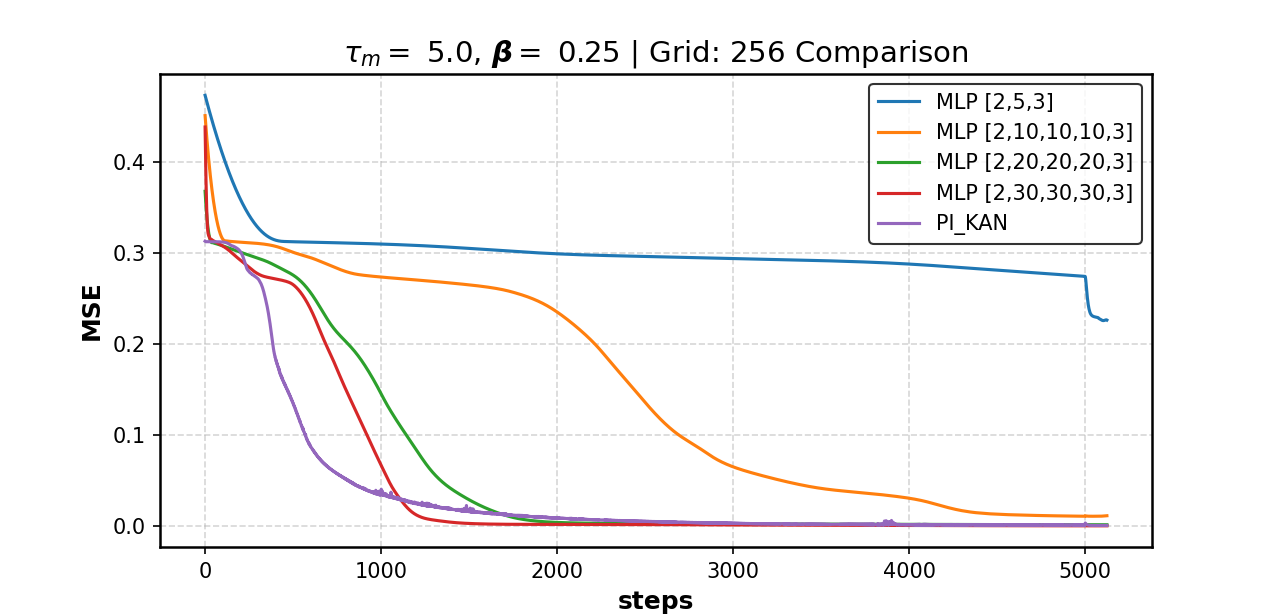}
    \caption{Comparative plot of  mean squared error (MSE) vs training iterations for $\tau=5.0$ and $\beta=0.25$ for various architectures.}
    \label{fig:13}
\end{figure*}

To evaluate the influence of neuron distribution on the training dynamics and predictive accuracy of the MultKAN framework, a comparative analysis was conducted under strictly equal parameter constraints. Seven distinct hidden-layer architectures were tested for the base case of $\tau_m = 1.0$ and $\beta = 1.0$: a baseline Straight-Uniform model, alongside U-shaped, Arch-shaped, V-shaped, Ascending, Inverted V, and Descending geometries. Table.~\ref{tab:architecture_comparison} presents a summary of the converged Mean Squared Error (MSE) and the individual physical component losses (PDE, Vorticity, BC, and Divergence) for each configuration. A global assessment of the table reveals that ``bottleneck" architectures—specifically the U-shaped and V-shaped models—achieve the lowest numerical MSE values ($1.80 \times 10^{-5}$ and $2.16 \times 10^{-5}$, respectively). However, the baseline Straight-Uniform model demonstrates highly competitive and stable performance, converging to an excellent error floor of $3.63 \times 10^{-5}$. Conversely, configurations that expand in the middle (Inverted V) or continuously widen (Ascending) struggled to minimize the error to the same degree, resulting in final MSEs an order of magnitude higher.

\begin{table*}[!htbp]
\centering
\resizebox{\textwidth}{!}{
\begin{tabular}{|c|l|c|c|c|c|c|c|}
\hline
\textbf{Sl no} & \multicolumn{1}{c|}{\textbf{Architecture}} & \textbf{Name} & \textbf{MSE Loss} & \textbf{PDE Loss} & \textbf{Vorticity Loss} & \textbf{BC Loss} & \textbf{Div Loss} \\ \hline
1 & [[2,0],[5,5],[5,5],[5,5],[5,5],[5,5],[3,0]] & \begin{tabular}[c]{@{}c@{}}Straight\\ Horizontal Line\end{tabular} & 3.63956e-05 & 4.84614e-01 & 3.14954e-04 & 1.20876e-04 & 3.98829e-04 \\ \hline
2 & [[2,0],[7,7],[4,4],[3,3],[4,4],[7,7],[3,0]] & \begin{tabular}[c]{@{}c@{}}U-shaped -\\ Valley\end{tabular} & 1.80244e-05 & 4.69187e-01 & 1.59301e-04 & 2.60768e-04 & 2.42350e-04 \\ \hline
3 & [[2,0],[3,3],[6,6],[7,7],[6,6],[3,3],[3,0]] & \begin{tabular}[c]{@{}c@{}}Arch\\ Shaped\end{tabular} & 9.75000e-05 & 4.81339e-01 & 7.12493e-04 & 4.33022e-04 & 1.62807e-03 \\ \hline
4 & [[2,0],[9,9],[2,2],[3,3],[2,2],[9,9],[3,0]] & \begin{tabular}[c]{@{}c@{}}V Shaped\end{tabular} & 2.16016e-05 & 4.76919e-01 & 2.05174e-04 & 2.67647e-04 & 2.97325e-04 \\ \hline
5 & [[2,0],[3,3],[4,4],[5,5],[6,6],[7,7],[3,0]] & \begin{tabular}[c]{@{}c@{}}/ -\\ Ascending\end{tabular} & 1.43453e-04 & 4.89940e-01 & 9.80241e-04 & 1.03037e-03 & 3.03062e-03 \\ \hline
6 & [[2,0],[2,2],[5,5],[11,11],[5,5],[2,2],[3,0]] & \begin{tabular}[c]{@{}c@{}}Inverted V\\ - Peak\end{tabular} & 1.00300e-04 & 4.88806e-01 & 8.74144e-04 & 3.76409e-04 & 1.12561e-03 \\ \hline
7 & [[2,0],[7,7],[6,6],[5,5],[4,4],[3,3],[3,0]] & \begin{tabular}[c]{@{}c@{}}\textbackslash -\\ Descending\end{tabular} & 2.41081e-05 & 4.68898e-01 & 2.46936e-04 & 2.95965e-04 & 2.51804e-04 \\ \hline
\end{tabular}%
}
\caption{Comparison of different MultKAN architectures with equal parameter constraints for $\tau=1$ and $\beta=1$}
\label{tab:architecture_comparison}
\end{table*}

Fig.~\ref{fig:13} illustrate the evolution of the mean squared error (MSE) and various Losses, respectively. \begin{equation}
\text{MSE} = \frac{1}{N} \sum_{i=1}^{N} \left[ \left( u_{pred}^{i} - u_{exact}^{i} \right)^2 + \left( v_{pred}^{i} - v_{exact}^{i} \right)^2 \right],
\label{eq:mse}
\end{equation}

where $N$ denotes the aggregate of spatial location points examined. The variables $(u_{pred}^{i}, v_{pred}^{i})$ and $(u_{exact}^{i}, v_{exact}^{i})$ represent the velocity vector predicted by the network and its analytical counterpart at each $i$-th coordinate, respectively. In Eq.(21), the "i" denotes the counter that iterates through every individual point in the grid to calculate the Mean Squared Error. The U-shaped and V-shaped architectures exhibit highly accelerated convergence rates, undergoing a steep reduction in error starting around iteration 100.  Also, to check the performance of the PI-KAN model, we tested it against widely used MLP models. The MLP architectures used are of two types- a single hidden-layer MLP (3 layer network) and a three hidden-layer MLP (5 layer network). The single hidden-layer MLP is constructed as [2,5,3] - one input layer with 2 nodes, one hidden layer with 5 nodes and one output layer with 3 nodes. Similarly the 3 hidden-layer MLP is constructed as one input layer with 2 nodes, one output layer with 3 nodes, the inner 3 hidden layers had varying number of nodes - one model had 10 nodes, one had 20 nodes and the last one had 30 nodes in each layer. Therefore, there are [2,10,10,10,3], [2,20,20,20,3] and [2,30,30,30,3]. In contrast, the PI-KAN had only 7 nodes in each of its 3 hidden layers, constructed as [2,7,7,7,3]. Accordingly, we see that the performance of the MLP model with 30 nodes, comes close to the performance of the PI-KAN with 7 nodes, but is inefficient when it comes to capturing even more intricate vortices. The MSE decreases more gradually in the PI-KAN model than in the MLP model.

\begin{figure*}[!htbp]  
    \centering
    \includegraphics[width=1.0\textwidth]{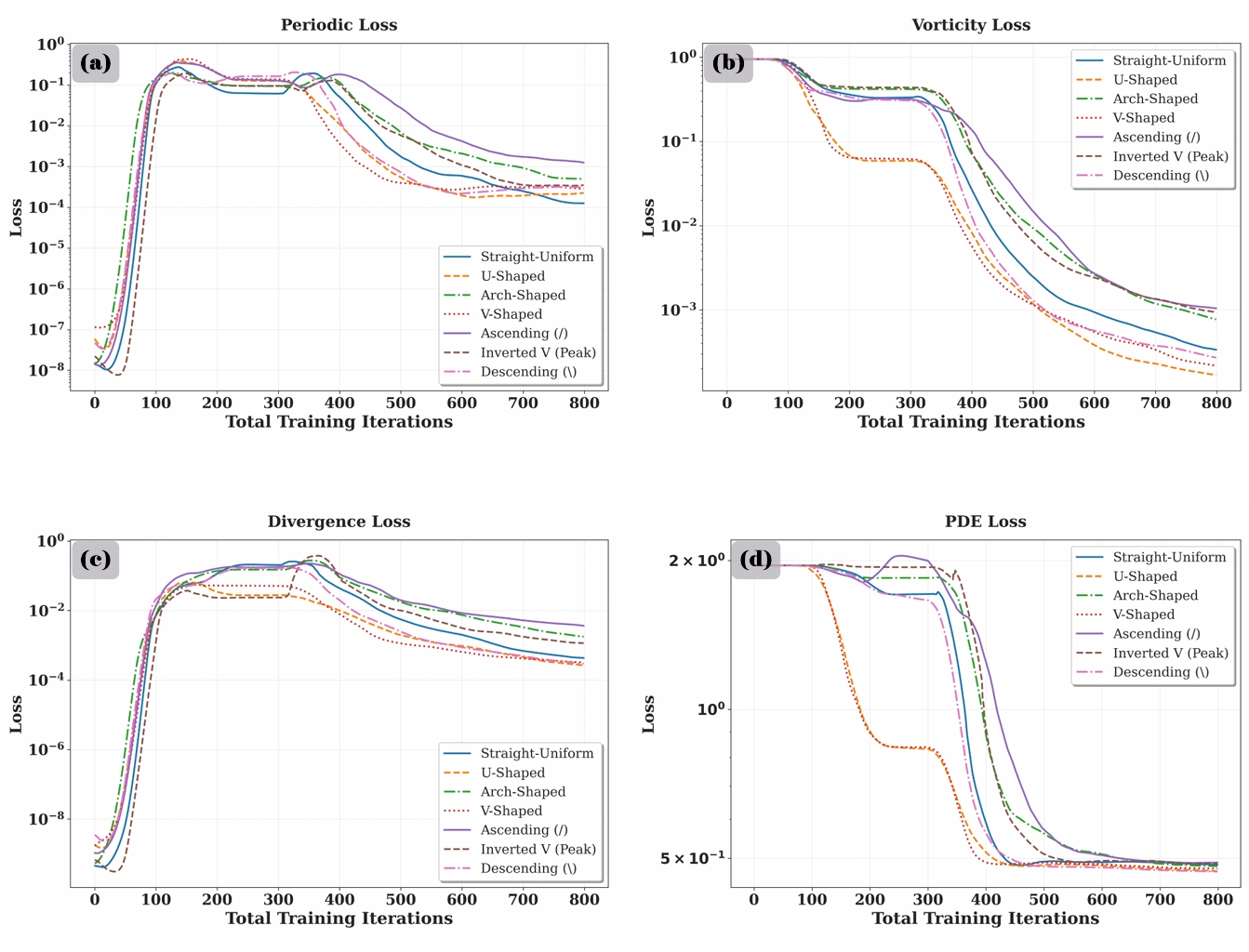}
        \caption{Semilogy plot of  mean squared error (MSE) vs training iterations for $\tau=1$ and $\beta=1$ for various architectures.}
    \label{fig:All_Losses_comparison}
\end{figure*}

Fig \ref{fig:All_Losses_comparison}  breaks down this optimization process, revealing that this rapid convergence is driven by their efficiency in minimizing the PDE and Vorticity residuals early in training. Forcing the fluid variables through a narrower set of hidden neurons acts as an implicit regularizer, coercing the network to learn the dominant, low-dimensional physical modes of the standard Taylor-Green vortex.

\begin{figure}[!h]
\centering
\includegraphics[width=0.5\textwidth]{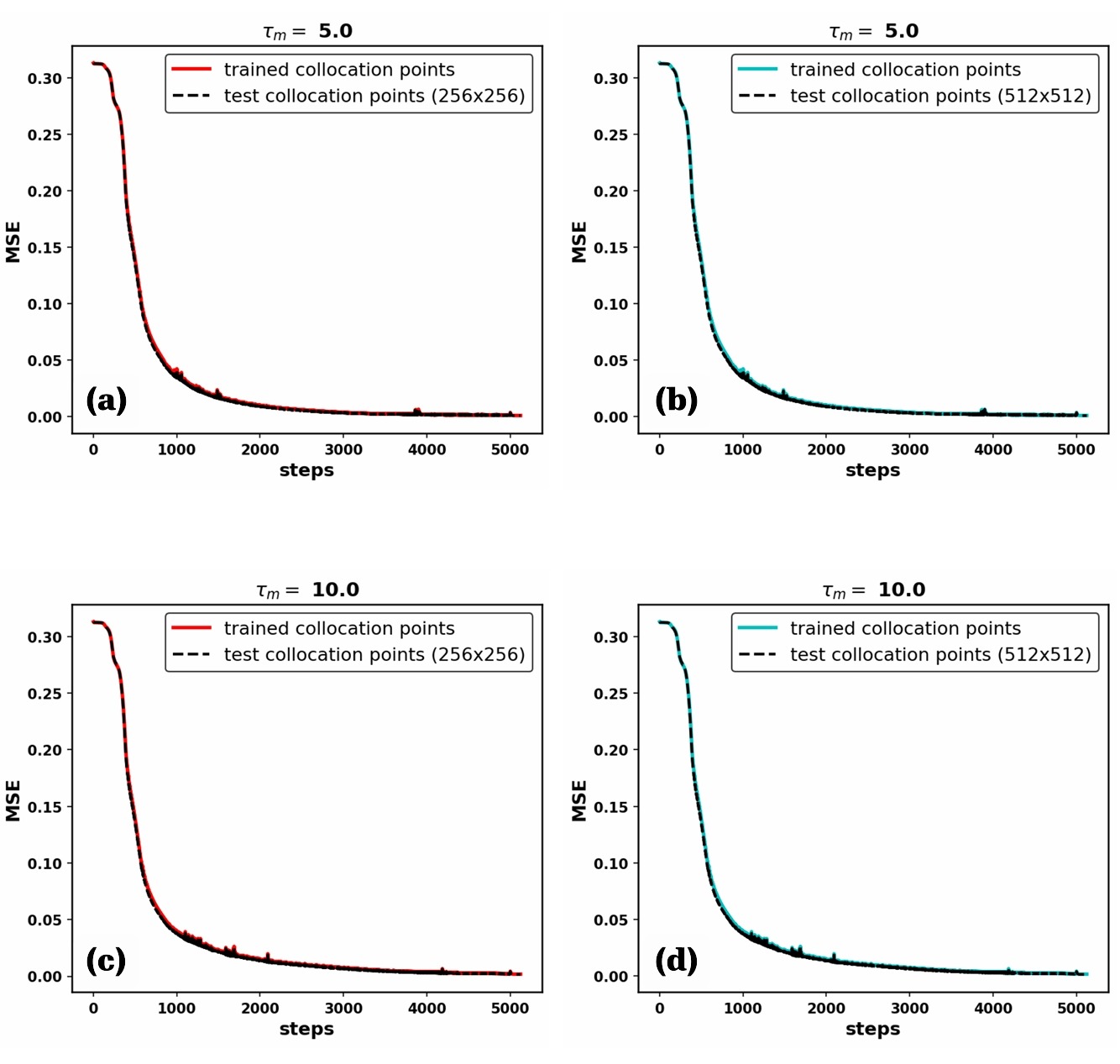}
\caption{The evolution of the Mean Squared Error (MSE) over the training steps on both trained and test collocation points. The MSE evolution curves for the MultKAN under the following conditions: (a) $\tau_m = 5.0$ evaluated on a $256 \times 256$ test grid, (b) $\tau_m = 5.0$ evaluated on a $512 \times 512$ test grid, (c) $\tau_m = 10.0$ evaluated on a $256 \times 256$ test grid, and (d) $\tau_m = 10.0$ evaluated on a $512 \times 512$ test grid.}
\label{fig:test&trained}
\end{figure}

After examining the effect of network architecture, we next evaluated the spatial generalization of the trained model. The choice of collocation points involves a trade-off between computational cost and physical accuracy. Dense grids generally improve the resolution of the flow structures. However, they also increase the computational expense significantly, while excessively sparse grids may cause the network to overfit the training coordinates instead of learning the underlying continuous flow behavior. In the present study, training was performed using a relatively sparse $41 \times 41$ collocation grid. To test whether the learned solution extends beyond these discrete training locations, the trained network was further evaluated on unseen grids of resolutions $256 \times 256$ and $512 \times 512$. Figure~\ref{fig:test&trained} presents the evolution of the Mean Squared Error (MSE) for $\tau_m = 5.0$ and $\tau_m = 10.0$. The test and training MSE curves remain closely aligned throughout the optimization process. This behavior indicates good spatial generalization. It also confirms that the sparse $41 \times 41$ grid is sufficient to recover the overall flow structures without a significant loss in predictive accuracy.

\begin{figure}[!h]
\centering
\includegraphics[width=0.45\textwidth]{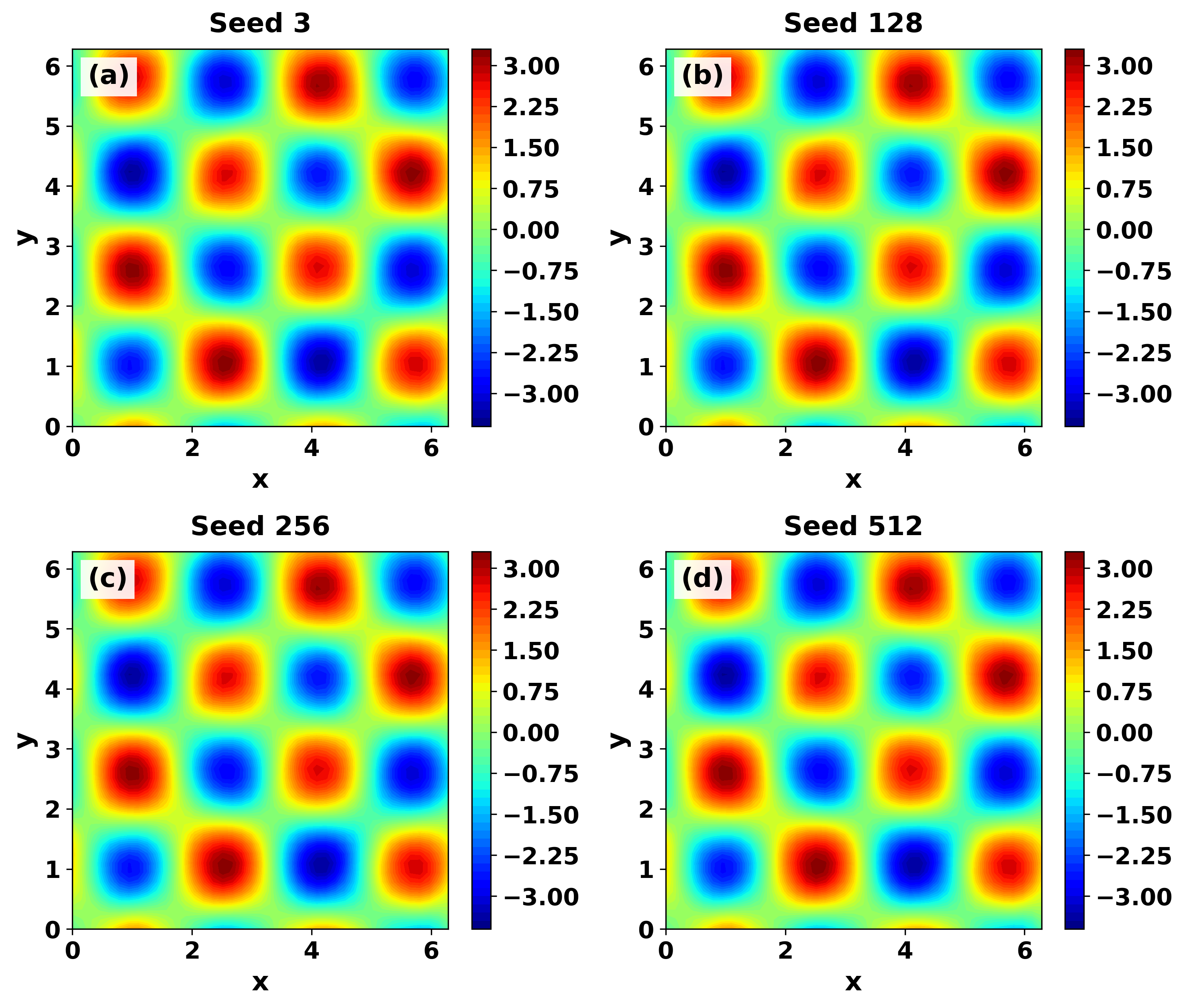}
\caption{The predicted vorticity fields generated by MultKAN architecture [2, 9, 9, 9, 9, 3] under different random seeds, demonstrating seed independency of architecture for $\tau_m = 20.0$ and $\beta = 0.25$. The predicted flow field results of the neural network for the cases of: (a) seed = 3, (b) seed = 128, (c) seed = 256, and (d) seed = 512.}
\label{fig:seed_independency}
\end{figure}

Beyond spatial discretization, neural network solvers frequently suffer from sensitivity to weight initialization, where different random seeds can yield diverging or unstable solutions. To rigorously test the robustness of our framework against this phenomenon, we investigated its seed independency under the most numerically stiff flow regime ($\tau_m = 20.0$ and $\beta = 0.25$). Figure \ref{fig:seed_independency} presents the predicted vorticity contours generated using four completely distinct random seeds (3, 128, 256, and 512). The predicted flow fields across all four initializations are visually and numerically indistinguishable. They consistently converge to the correct physical state, maintaining well-aligned isoclines and matching the exact rotational magnitudes of the reference structures. This uniformity clearly demonstrates that the proposed PI-KAN architecture, when coupled with gradient clipping and dynamic loss weighting, effectively eliminates initialization bias and guarantees reproducible predictions even for highly nonlinear viscoelastic fluids.

In the last section, we summarize our results. 

\section{Conclusions}
\label{Sec:Conclusion}

This work introduces systematic investigation of the employment of the architecture of Physics-Informed Kolmogorov--Arnold Networks (PI-KANs) to study the solution of viscoelastic fluid equations. 
Building upon the earlier foundations of PINNs  and their advanced variants, this study aims to:
(i) evaluate and understand the representational capacity of KANs in capturing nonlinear constitutive behavior in viscoelastic flow, 
(ii) analyze their convergence and training efficiency in comparison to the standard PINN architectures under different flow regimes, and 
(iii) explore their generalization across geometries, including benchmark cases involving obstacle-induced viscoelastic wakes. 
The proposed PI-KAN framework bridges neural-operator theory and physics-informed modeling, providing new insights into data-driven simulation of complex rheological flows. The PI-KAN model exhibits a more gradual reduction in MSE than the MLP model, indicating improved training stability and robustness, thereby making it a more reliable approach for solving complex physics-based problems.
Despite the slight convergence speed advantage of the bottleneck geometries in this baseline scenario, the Straight-Uniform architecture was strategically selected as the primary configuration for the broader parametric study. While a bottleneck is efficient for $\tau_m = 1.0$, simulating highly viscoelastic flows ($\tau_m = 10.0$ and $20.0$) introduces severe mathematical stiffness and sharp, high-frequency stress gradients. Maintaining a uniform, high capacity across all hidden layers ensures that the network does not overly compress critical spatial information, providing the robust degrees of freedom necessary to resolve complex viscoelastic instabilities without introducing artificial bottlenecks.
\section*{Data and code availability}
Data from this study and the computer scripts can be obtained from the authors upon 
reasonable request.
\section*{Conflicts of Interest}
No conflicts of interest, financial or otherwise, are declared by the authors.
\section*{Author Contributions} 
SS, MS, and AG planned the research; SS and MS carried out the calculations and analysed the numerical data; SS and MS prepared the tables, figures, and the draft of the manuscript; SS, MS and AG then revised the manuscript in detail and approved the final version.
\section*{Acknowledgments}
SS acknowledges the internship opportunity at the Department of Physics, Maulana Azad National Institute of Technology (MANIT) Bhopal, India. MS acknowledges the UGC National Scholarship for Post Graduate Studies for financial support. The authors thank the Central Computing Resources at MANIT Bhopal, India, for providing computational support.
\bibliographystyle{apsrev4-2}
\bibliography{paper_bib}
\end{document}